\documentclass[a4paper,11pt]{article}
\pdfoutput=1 

\usepackage{jheppub} 
\usepackage[T1]{fontenc} 

\usepackage{epsf}
\usepackage{amssymb}
\usepackage{amsmath}
\usepackage{amsbsy} 
\usepackage{fancybox}
\usepackage{color}
\usepackage{hyperref}
\allowdisplaybreaks

\usepackage{graphicx,epsfig,url}
\usepackage{floatrow}
\let\al=\alpha
\let\be=\beta

\let\Ga=\Gamma

\let\la=\lambda

\let\th=\theta

\let\om=\omega

\def\lambdabar{\protect\@lambdabar}
\def\@lambdabar{%
\relax
\bgroup
\def\@tempa{\hbox{\raise.73\ht0
\hbox to1pt{\kern.25\wd0\vrule width.7\wd0
height0.5pt depth.1pt\hss}\box0}}%
\mathchoice{\setbox0\hbox{$\displaystyle\lambda$}\@tempa}%
{\setbox0\hbox{$\textstyle\lambda$}\@tempa}%
{\setbox0\hbox{$\scriptstyle\lambda$}\@tempa}%
{\setbox0\hbox{$\scriptscriptstyle\lambda$}\@tempa}%
\egroup
}

{\setlength{\fboxsep}{15pt}
\setlength{\mylength}{\linewidth}%
\addtolength{\mylength}{-2\fboxsep}%
\addtolength{\mylength}{-2\fboxrule}%
\Sbox
\minipage{\mylength}%
\setlength{\abovedisplayskip}{0pt}%
\setlength{\belowdisplayskip}{0pt}%
\equation}%
{\endequation\endminipage\endSbox \[\fbox{\TheSbox}\]}

\def\beq{\begin{equation}}
\def\eeq{\end{equation}}
\def\bea{\begin{eqnarray}}
\def\eea{\end{eqnarray}}
\def\bitem{\begin{itemize}}
\def\eitem{\end{itemize}}
\newcommand{\bec}{\begin{center}}
\newcommand{\eec}{\end{center}}
\newcommand{\ba}{\begin{array}}
\newcommand{\ea}{\end{array}}

\renewcommand\vec[1]{\ensuremath\boldsymbol{#1}}

\def\bo{{\raise.15ex\hbox{\large$\Box$}}}               
\def\face{{\raise.2ex\hbox{$\displaystyle \bigodot$}\mskip-2.2mu \llap {$\ddot
        \smile$}}}                                      

\def\abs#1{\left| #1\right|}                    
\def\leftrightarrowfill{$\mathsurround=0pt \mathord\leftarrow \mkern-6mu
        \cleaders\hbox{$\mkern-2mu \mathord- \mkern-2mu$}\hfill
        \mkern-6mu \mathord\rightarrow$}       
\def\dvec#1{\vbox{\ialign{##\crcr
        \leftrightarrowfill\crcr\noalign{\kern-1pt\nointerlineskip}
        $\hfil\displaystyle{#1}\hfil$\crcr}}}           

\definecolor{darkred}{rgb}{0.8,0,0}
\definecolor{darkblue}{rgb}{0,0,0.7}

\usepackage{amsmath}	
\usepackage{amssymb}
\usepackage{multirow}
\usepackage{siunitx}
\usepackage{physics}
\usepackage{arydshln}

\newcommand{\bigzero}{\mbox{\normalfont\Large 0}}
\renewcommand{\vec}[1]{\mathbf{#1}}
\newcommand{\Lagr}{\mathcal{L}}

\author{Dugald Hepburn}
\author{and Stephen M. West}
\affiliation{Department of Physics, Royal Holloway, University of London, \\ Egham, Surrey, TW20 0EX, United Kingdom}

\title{\boldmath{Dark Matter and Neutrino Masses in a Portalino-like Model}}

\emailAdd{Dugald.Hepburn@rhul.ac.uk}
\emailAdd{Stephen.West@rhul.ac.uk}

\abstract{
We explore a Portalino-like model of dark matter and neutrino masses in which right-handed neutrino fields connect gauge neutral operators from the Standard Model and Hidden Sector. Neutrino masses are generated via a seesaw-like mechanism that can explain the light active neutrino masses. The model includes a ``Portalino'' state that connects the two sectors via the neutrino portal. Dark Matter in this model consists of a hidden sector Dirac fermion that dominantly freezes-out via resonant annihilations into other hidden sector states, which ultimately results in a population of Portalinos. Due to small mixing in the extended neutrino sector these Portalinos tend to be cosmologically long lived, decaying into Standard Model particles leading to constraints on the model from Big Bang Nucleosynthesis and measurements of the Cosmic Microwave Background radiation. Combining these limits with direct constraints on the size of the Portalino-neutrino mixing and the assumptions of the model the viable mass ranges for the Portalino states are found to be $\SI{0.02}{\electronvolt}\lesssim m_n \lesssim \SI{6.4}{\electronvolt}$ or $\SI{489}{\mega\electronvolt} \lesssim m_n \lesssim$ TeV. Indirect dark matter signals in the form of highly boosted, mono-energetic Portalinos produced in Dark Matter annihilations provide a target for neutrino telescopes.}

\begin{document} 
\maketitle
\flushbottom


\section{Introduction}
\label{section:introduction}

The question of how the dark sector interacts with the visible sector, if it does at all, underpins the uncertainty surrounding the nature and origin of Dark Matter (DM). Many proposals have been made for how a connection can be established through so called ``portals'', including the Higgs portal see e.g. \cite{Schabinger:2005ei, Patt:2006fw,MarchRussell:2008yu}, through the Kinetic mixing portal \cite{Holdom:1985ag,Dienes:1996zr,DelAguila:1993px,Babu:1996vt}, neutrino portal \cite{Lindner:2010rr,Falkowski:2009yz,GonzalezMacias:2015rxl,Batell:2017rol,Batell:2017cmf,Blennow:2019fhy}, axion portal \cite{Nomura:2008ru}, or perhaps there is no portal at all in which case the dark sector evolves independently but may still have observable effects \cite{March-Russell:2020nun}.

In this paper we focus on the neutrino portal, and in particular examine a model inspired by the Portalino scenario in which a singlet fermion field connects gauge neutral fermion operators from the Standard Model (SM) and hidden sector \cite{Schmaltz:2017oov}.  

In a simple realisation of the Portalino framework introduced in \cite{Schmaltz:2017oov} the SM is supplemented by two additional gauge singlet fermions and a complex scalar singlet. One of the fermion states plays the role of the right-handed neutrino, $\nu_R$, and couples to the gauge invariant combination of the SM Higgs and Lepton doublets generating a Dirac like neutrino mass term after electroweak symmetry breaking. This right-handed neutrino state also couples to a second gauge invariant operator composed of the second singlet fermion, which we call $\psi$, and the complex scalar field, call it $\Phi$. If there is a dark $U(1)$ under which $\Phi$ and $\psi$ both transform then we can construct Yukawa interactions that lead to Dirac masses after the spontaneous symmetry breaking of the dark $U(1)$ such that 

\bea\label{eq:origlag}
\Lagr \subset \la_1 \nu_R^{\dagger}H_0 \nu_L+ \la_2 \nu_R^{\dagger} \Phi \;\psi=m_d \nu_R^{\dagger} (\sin \theta \;\nu_L+ \cos \theta  \; \psi),
\eea

where $H_0$ is the neutral component of the SM Higgs doublet, and the linear combination of $\psi$ and $\nu_L$ forms a massive Dirac state with $\nu_R$. As $\psi$ has vector interactions with the dark gauge sector the light neutrino (zero) mass eigenstate, $\nu = \cos \theta \;\nu_L- \sin \theta  \; \psi$ inherits these interactions albeit suppressed by a factor of $\sin \theta$. As pointed out in \cite{Schmaltz:2017oov}, this scenario is a specific version of a $Z^{\prime}$ model in which the only interactions between the new hidden sector $U(1)$ and the SM is via the neutrinos.\footnote{Although, given the introduction of a new scalar field, the Higgs portal also connects the two sectors.}

Introducing DM into the hidden sector is straightforward. For example, in \cite{Schmaltz:2017oov} a Yukawa interaction involving the scalar state $\Phi$ and a new Dirac fermion, call it $X$, was included. The dark sector dominantly interacts with the neutrino sector potentially leading to the $X$ DM states freezing-out via annihilation to neutrinos. This removes, or greatly suppresses, the usual modes for probing DM in direct and indirect detection detection experiments allowing for models that consider a wider range of potentially viable DM masses. On the other hand this makes the model harder to probe. 

The simple model outlined above however requires modification in order to include neutrino masses. There are a number of choices we can make to do this. One possibility is to add a Majorana mass term for the $\psi$ field leading to a model along the lines of the inverse see-saw model, see e.g. \cite{Dias:2012xp}.  In \cite{Liu:2019ixm}, it was suggested that it may be possible to produce non-zero neutrino masses in a Zee-type model including two Higgs doublets via a $(l_i h)(h l_j)$ term generated at loop level.

An alternative is to change particle content by introducing further generations of the singlet fermion fields. In this paper, we extend the model to include two more generations of right-handed neutrino and introduce associated large Majorana mass terms for these states. This set-up generates masses for two of the three generations of light neutrinos, with the heavy Majorana masses suppressing the mass scale of these two mass eigenstates through a seesaw-like mechanism. Without the heavy Majorana masses, the light neutrinos will be Dirac states with Dirac neutrino masses of $\mathcal{O} \left(\lambda_{\nu} v_h\right)$. Although with sufficiently small Yukawa couplings this is in principle a viable model, we choose instead to adopt the Majorana case.

The introduction of the large Majorana mass scale leads to small mixing angles in this combined neutrino-hidden state sector, which in turn generates suppressed couplings for the more massive hidden sector states. This leads to relatively long lifetimes for these states, giving rise to interesting cosmological implications and constraints on the model.
 
In Section~\ref{section:model} we describe the model in full, including the detailed properties of the putative DM candidate. In Section~\ref{section:neutrino} we outline the model's predictions for neutrino masses and mixings, and how the experimentally observed values can be accommodated. In Section~\ref{section:abundance} we specify the viable parameter space capable of generating the correct DM abundance. We explore the phenomenology of - and constraints on - the new hidden sector states, which can have lifetimes up to and exceeding the age of the universe, in Section~\ref{section:pheno}. 

\section{The Model}
\label{section:model}

The model consists of the SM supplemented by a number of SM singlet fields. These include three generations of right-handed neutrino, $\nu_{R_{\al}}$ ($\al=$ 1, 2, 3),  a complex scalar, $\Phi$, and three Weyl fermions, $\psi$, $X_L$ and $X_R$. The $X_L$, $X_R$ fields will combine to form a Dirac fermion state and will be our DM candidate. We further introduce a new abelian gauge symmetry, $U(1)_{\rm d}$, under which $\Phi$, $\psi$ and $X_R$ transform each with charge $1/2$.  The right-handed neutrinos, $X_L$ and all other SM states are uncharged under the new symmetry. Additionally both $X_L$ and $X_R$ are charged under a separate $Z_3$ symmetry uncharged. The role of this $Z_3$ is two-fold, firstly this forbids an explicit Majorana mass term for $X_L$ and secondly it ensures the stability of the $X$ DM state.  A summary of these charges is displayed in Table~\ref{tab:charges}.

\begin{table}[t]
\caption{Charge assignments of the field content in the hidden sector under $U(1)_{\rm d}$ and $Z_3$. All fields in the table are Standard Model singlets}
\begin{center}
\begin{tabular}{|c|c|c|c|c|c|}\hline
Field  &  $\nu_{R_\al}$ & $\Phi$ & $\psi$ & $X_L$ & $X_R$ \\
\hline
$U(1)_{\rm d}$  &  0  & 1/2 & 1/2 & 0 & 1/2 \\
 $Z_3$ & + & + & + & $e^{i\pi/3}$ & $e^{i\pi/3}$\\
\hline
\end{tabular}
\end{center}
\label{tab:charges}
\end{table}%

Given this particle content and charge assignment, the Lagrangian for the model reads

\bea\nonumber
\Lagr = \left(- \sqrt{2} \lambda^{\nu}_{\al\be}L^{\dagger}_{\al}H\nu_{R\beta} -\sqrt{2} \lambda^{\psi}_\alpha \psi^{\dagger} \Phi \nu_{R\alpha}  + \frac{i}{2} M_{R\alpha \beta} \nu_{R\alpha}^{T} \sigma_2 \nu_{R\beta} -\sqrt{2} \lambda_{X} X_R^{\dagger} \Phi X_L + {\rm h.c.} \right)\\\nonumber +\mu_H^2 |H|^2 - \lambda_H |H|^4 + \mu_{\Phi}^2 |\Phi|^2 - \lambda_{\Phi} |\Phi|^4 - \lambda_{H,\Phi} |H|^2 |\Phi|^2 \ldots,\\
\eea
where the ellipsis represents the SM Lagrangian terms  and all kinetic terms for the new states including all relevant gauge interactions with the $U(1)_{\rm d}$ gauge boson, $\om_{\mu}$, and we specify that $\mu_{\Phi}^2, \mu_H^2>0$. In principle, we may expect a kinetic mixing term that mixes the field strengths associated with the new $U(1)_{\rm d}$ and SM hypercharge $U(1)_Y$. However, we assume, for simplicity, that this term is sufficiently small that it does not impact the phenomenology of the model\footnote{Following \cite{Gherghetta:2019coi}, the leading contribution to the loop induced kinetic mixing term is $\sim 10^{-7} \left(\lambda^{\nu}\lambda^{\psi}\right)^2$, which arises from a 3-loop diagram. Even with the couplings $\lambda^{\nu}$ and $\lambda^{\psi}$ of order 1 the size of the induced kinetic mixing term is not relevant for the model phenomenology and evades bounds on the size of the kinetic mixing parameter, \cite{Gherghetta:2019coi}.}.

The form of the potential leads to the spontaneous breaking of $SU(2)_{\rm L}\times U(1)_{\rm y}\rightarrow U(1)_{\rm em}$ and $U(1)_{\rm d} \rightarrow$  nothing. We parameterise both $\Phi$ and $H$ in terms of excitations, $\phi^{\prime}$ and $h^{\prime}$ respectively, around the corresponding vacuum expectation values, expressed in the unitary gauge as $H = \frac{1}{\sqrt{2}} \begin{pmatrix}
0 \\
v_h + h'
\end{pmatrix} $ and $ \Phi = \frac{1}{\sqrt{2}} \left(v_{\phi} + \phi'\right)$, where the expectation values are given by

\begin{align}
v_h^2 = \frac{2 \mu_H^2 \lambda_{\Phi} -  \mu_{\Phi}^2 \lambda_{H, \Phi}  }{4 \lambda_H \lambda_{\Phi} - \lambda_{H,\Phi}^2}, \qquad v_{\phi}^2 = \frac{2 \mu_{\Phi}^2 \lambda_H -  \mu_{H}^2 \lambda_{H, \Phi}  }{4 \lambda_{\Phi} \lambda_{H} - \lambda_{H,\Phi}^2}.
\end{align}

The Lagrangian after spontaneous symmetry breaking reads

\begin{align}\label{eq:lagflav}
&\Lagr =\left(-M_{d\alpha \beta} \nu_{l\alpha}^{\dagger} \nu_{R\beta} -M_{\psi_{\alpha}} \psi^{\dagger} \nu_{R\alpha} + \frac{i}{2} M_{R\alpha \beta} \nu_{R\alpha}^{T} \sigma_2 \nu_{R\beta} - m_{X} 
X_R^{\dagger} X_L + {\rm h.c.}\right)\nonumber\\
&+\left(-\lambda^{\nu}_{\alpha \beta} \nu_{l\alpha}^{\dagger} \nu_{R\beta} h'  - \lambda^{\psi}_{\alpha} \psi^{\dagger} \phi' \nu_{R\alpha} - \lambda_{X} X_R^{\dagger} \phi' X_L + {\rm h.c.}\right)-V(h^{\prime},\phi^{\prime})+\ldots, 
\end{align}
where $M_{d\alpha \beta} = \lambda^{\nu}_{\alpha \beta} v_{h}, \;\;M_{\psi\alpha} = \lambda^{\psi}_{\alpha} v_{\phi},\;\; m_{X} = \lambda_{X} v_{\phi}$,  and where we have assumed $\lambda_X$ is real. The scalar potential now reads
\begin{align}
&V(h^{\prime},\phi^{\prime}) = \lambda_H v_h^2 h'^2 + \lambda_{\Phi} v_{\phi}^2 \phi'^2 + \lambda_{H,\Phi} v_h v_{\phi} h' \phi'  \nonumber
\\
&\qquad \qquad \qquad + \lambda_H v_h h'^3 + \frac{\lambda_H}{4} h'^4 + \lambda_{\Phi} v_{\phi} \phi'^3 + \frac{\lambda_{\Phi}}{4} \phi'^4  \nonumber
\\
&\qquad \qquad \qquad + \frac{\lambda_{H,\Phi} v_h}{2}  h' \phi'^2 + \frac{\lambda_{H,\Phi} v_{\phi}}{2} h'^2 \phi' + \frac{\lambda_{H,\Phi}}{4} h'^2 \phi'^2.
\end{align}

The ellipsis in Equation~\ref{eq:lagflav} again include the rest of the SM Lagrangian with the addition of all the beyond the SM kinetic terms and interactions of the states charged under $U(1)_{\rm d}$ with the associated gauge boson, $\om$, whose mass is given by $m_{\om}=(v_{\phi} \tilde{g})/2$ after symmetry breaking.

The scalar sector is diagonalised via the rotation defined by
\begin{align}
\label{eq:scalarmix}
\begin{pmatrix}
h \\
\phi
\end{pmatrix}
=
\begin{pmatrix}
\cos \theta & \sin \theta \\
-\sin \theta & \cos \theta
\end{pmatrix}
\begin{pmatrix}
h' \\
\phi'
\end{pmatrix}, \qquad \mbox{where} \qquad \tan 2 \theta = - \frac{\lambda_{H,\Phi}v_hv_{\phi}}{\lambda_{\Phi}v_{\phi}^2 - \lambda_{H}v_h^2}.
\end{align}
\noindent
The measured value of the couplings of the SM gauge bosons to the Higgs are very close to that predicted by the SM and consequently the mixing angle $\theta$ must be small - in the region of interest ($v_{\phi} \gtrsim \si{\tera\electronvolt}$), the limit is approximated by \cite{Lebedev:2021xey,Falkowski:2015iwa}:

\begin{align}
    |\sin \theta| \lesssim \frac{0.3}{\sqrt{1 + \log\left( \frac{m_{\phi}}{\si{\tera\electronvolt}}\right) }}.
\end{align}

This can be achieved by insisting $v_{\phi} \gg v_h$ and by assuming that the coupling $\la_{H,\phi}$ is moderately suppressed compared with the other dimensionless couplings in the scalar potential. Suppressing $\la_{H,\phi}$ also has the effect of shutting off the Higgs Portal as a channel for DM annihilation, see Section~\ref{section:abundance} for details. The $v_{\phi} \gg v_h$ hierarchy is also necessary for achieving light neutrinos with phenomenologically viable masses. 

In this limit the scalar mass eigenstates read
\begin{align}
m_{h}^2 &= 2 \lambda_H v_h^2 \left( 1 - \frac{\lambda_{H, \Phi}^2}{4\lambda_{\Phi}\lambda_{H}} +\mathcal{O}\left( \left(\frac{\lambda_{H,\Phi} v_h}{\lambda_{\Phi}v_{\phi}}\right)^2\right) \right),
\\
m_{\phi}^2 &= 2 \lambda_{\Phi} v_{\phi}^2 \left( 1 + \left(\frac{\lambda_{H,\Phi} v_h}{\lambda_{\Phi}v_{\phi}}\right)^2 + \mathcal{O} \left(\left(\frac{\lambda_{H,\Phi} v_h}{\lambda_{\Phi}v_{\phi}}\right)^2\frac{\lambda_H v_h^2}{\lambda_{\Phi} v_{\phi}^2} \right) \right).
\end{align}

Moving to the fermionic content of the model, the first two mass mixing terms of Equation~\ref{eq:lagflav} encode the Portalino-like mixing, as detailed in Equation~\ref{eq:origlag}. The picture is necessarily complicated by the Majorana mass term for the $\nu_R$ fields and this is what leads to non-zero light neutrino masses. In this work we do not propose a full flavour model, instead we assume that there are no significant hierarchies within the components of $\lambda^{\nu}_{\al\beta}$, $\lambda^{\psi}_{\al}$ or $M_{R\al}$. Under this assumption, we define
\bea\label{eq:flavour_defs}
\lambda^{\nu}_{\al\beta} \equiv \lambda^{\nu}F^{\nu}_{\al\beta}\equiv\frac{m_d}{v_u}F^{\nu}_{\al\beta},\;\; \lambda^{\psi}_{\al} \equiv \lambda^{\psi} F^{\psi}_{\al}\equiv \frac{m_{\psi}}{v_{\phi}} F^{\psi}_{\al},\;\;M_{R\al\be}\equiv m_R F_{R\al\be}
\eea where the parameters without flavour indices, which we define to be real, will be used to indicate the typical size of the entries of each term leaving the precise flavour dependence to the objects labelled $F$. 

In order to obtain the correct mass spectrum, we require that the $\nu_R$ Majorana mass is much larger than its mass mixing with either $\psi$ or the active neutrinos $\nu_L$, and that the mixing with $\psi$ is much larger than the mixing with $\nu_L$, that is $\lambda^{\nu} v_h \ll \lambda^{\psi} v_{\phi} \ll m_R$ or equivalently, $m_d\ll m_{\psi}\ll m_R$.

Given this hierarchy of scales, the mass matrix mixing the states $\psi, \nu_l$ and $\nu_R$ can be approximately diagonalised via the following transformations

\begin{align}
\label{eqn:mixings}
\nu_i & \sim U_{i \alpha}^{(\nu\nu_l)} \begin{pmatrix} \nu_{l \alpha} \\ i \sigma_2 \nu_{l \alpha}^* \end{pmatrix} 
+ \frac{m_d}{m_{\psi}}\; U_i^{(\nu\psi)} \begin{pmatrix} \psi \\ i \sigma_2 \psi^* \end{pmatrix} 
+ \frac{m_d}{m_R}\; U_{i \alpha}^{(\nu\nu_R)} \begin{pmatrix} -i \sigma_2 \nu_{R_{\alpha}}^* \\ \nu_{R_{\alpha}} \end{pmatrix}, \nonumber
\\
n &\sim \frac{m_d}{m_{\psi}} U_{\alpha}^{(n \nu_l)} \begin{pmatrix} \nu_{l_{\alpha}} \\ i \sigma_2 \nu_{l_{\alpha}}^* \end{pmatrix} 
+ U^{(n\psi)} \begin{pmatrix} \psi \\  i \sigma_2 \psi^* \end{pmatrix} 
+ \frac{m_{\psi}}{m_R} U_{\alpha}^{(n \nu_R)}\; \begin{pmatrix} -i \sigma_2 \nu_{R_{\alpha}}^* \\  \nu_{R_{\alpha}} \end{pmatrix}, \nonumber
\\
N_i &\sim \frac{m_d}{m_R} U_{i \alpha}^{(N\nu_l)} \begin{pmatrix} \nu_{l_{\alpha}} \\ i \sigma_2 \nu_{l_{\alpha}}^* \end{pmatrix} 
+ \frac{m_{\psi}}{m_R}\; U_i^{(N\psi)} \begin{pmatrix}  \psi \\  i \sigma_2 \psi^* \end{pmatrix} + U_{i \alpha}^{(N \nu_R)} \begin{pmatrix}  -i \sigma_2 \nu_{R_{\alpha}}^* \\ \nu_{R_{\alpha}} \end{pmatrix},
\end{align}
where $i, \al=1,2,3$ and in the above the definitions in Equation~\ref{eq:flavour_defs} have been used to factor out the leading order behaviour while the various factors of $U$ contain all the detailed flavour mixing. 

The full diagonalisation of the $(\nu_{l\alpha}, \psi, \nu_{R\alpha})$ system is presented in Appendix~\ref{section:mixing} and includes the full expression for the unitary matrix that Equation~\ref{eqn:mixings} derives from, including the explicit form of the $U$ factors. 

To leading order the three light neutrinos, $\nu_i$, have masses 
\begin{align} \label{eq:lightneumass}
m_{\nu_1} = 0, \quad m_{\nu_{2,3}} \sim \frac{m_{d}^2}{ m_R},
\end{align}
and the three heavy neutrinos, $N_i$, have masses \begin{align}
m_{N_i} \sim m_{R}.
\end{align}

The field $n$, which we choose to call the Portalino\footnote{It is not entirely clear which of our states is the analogue of the Portalino from the earlier example outlined in Equation~\ref{eq:origlag}, where $\nu_R$ was the Portalino. It should perhaps, morally speaking, be the fields $N_i$ that should take on the Portlino title given that their largest component comes from the $\nu_R$ fields. We prefer however to adopt the naming conventions from neutrino mass models where the $N_i$s are the heavy neutrinos, the $\nu_i$ the light neutrinos, leaving the $n$ field which we will refer to as the Portalino.}, has a mass suppressed relative to the mass scale $m_{\psi}$ given by

\begin{align}\label{eq:portmass}
m_n \sim \frac{m_{\psi}^2}{m_{R}}.
\end{align}

With $m_{\psi}=\la_{\psi}v_{\phi}$, it is clear that under the hierarchy assumption of $m_R\gg \lambda_{\psi} v_{\phi}$, it must be true that $m_n\ll v_{\phi}$. The full Lagrangian in the mass eigenbasis is given in Appendix~\ref{section:lagrangian}.

\section{Reconstructing the PMNS matrix}
\label{section:neutrino}

In Appendix~\ref{section:mixing} the full masses and mixings of the $(\nu_{l\alpha}, \psi, \nu_{R\alpha})$ system are calculated and presented as approximate analytic expressions following the hierarchy in masses scales $m_{d}\ll m_{\psi} \ll m_{R}$. As stated above, we are assuming that there are no significant hierarchies between the flavours of the individual masses. 

Due to the additional states mixing with the left-handed neutrinos, the PMNS matrix will no longer be unitary. Assuming no mixing in the charged lepton sector the PMNS matrix is  determined by the mixing in the extended neutrino sector only. The allowed 3$\sigma$ ranges on the entries PMNS matrix (once the assumption of unitary is dropped) are \cite{Parke:2015goa}:
\begin{align}
|V| = \begin{pmatrix}
0.76 \rightarrow 0.85 & 0.50 \rightarrow 0.60 & 0.13 \rightarrow 0.16 \\
0.21 \rightarrow 0.54 & 0.42 \rightarrow 0.70 & 0.61 \rightarrow 0.79 \\
0.18 \rightarrow 0.58 & 0.38 \rightarrow 0.72 & 0.40 \rightarrow 0.78 
\end{pmatrix}.
\end{align}

\noindent In addition to the mixing, the masses of the light neutrinos must fall within the following ranges (the lightest neutrino is massless in this model) - assuming normal ordering \cite{Esteban:2020cvm}:

\begin{align}
m_2 \in \left[ \SI{8.2}{\milli\electronvolt}, \SI{9.0}{\milli\electronvolt} \right], m_3 \in \left[ \SI{49.0}{\milli\electronvolt}, \SI{50.9}{\milli\electronvolt} \right].
\end{align}

A flavour model for the structure of $\lambda^{\nu}, \lambda^{\psi}$ and $M_R$ is beyond the scope of this work, and without such a model the task of finding values for the components of these matrices that satisfy the mixing and mass constraints is an under-constrained problem.

\section{Dark Matter Phenomenology}
\subsection{Dark Matter Abundance}
\label{section:abundance}

Moving now to the DM phenomenology in this model. In our numerical analysis below we have used micrOMEGAs, \cite{Belanger:2018ccd}, to compute the freeze-out abundance for a range of parameter values. We can eliminate a number of parameters in favour of the measured values of the Higgs mass, $m_h$, and the masses of the SM gauge bosons. The DM phenomenology is not sensitive to the relative sizes of the neutrino masses and mixings. In order to ensure that we consider parameter values that can reproduce light neutrino masses we set\footnote{This mass sets the scale for the light neutrinos, the precise masses and mixings are determined by other flavour parameters that do not play a leading role in the determination of the DM abundance. In order to numerically calculate the DM abundance we do need to input some structure by hand and we assume a simple parameterised form of the components of the full $7\times7$ neutrino mixing matrix, these are listed in Appendix~\ref{section:mixing_micro}.} $(\frac{m_d}{m_{\psi}})^2 m_n = \SI{20}{\milli\electronvolt}$ in order to fix $\la_{\nu}$.
The remaining relevant masses and couplings are determined, at least to leading order, by seven parameters: $v_{\phi}, \lambda_{\psi}, \lambda_{\Phi}, \lambda_{H,\Phi}, m_n$, $\tilde{g}$, and $m_{X}$. 

\begin{figure}[t]
\centering
\begin{minipage}{0.32\textwidth}
\includegraphics[width=\textwidth]{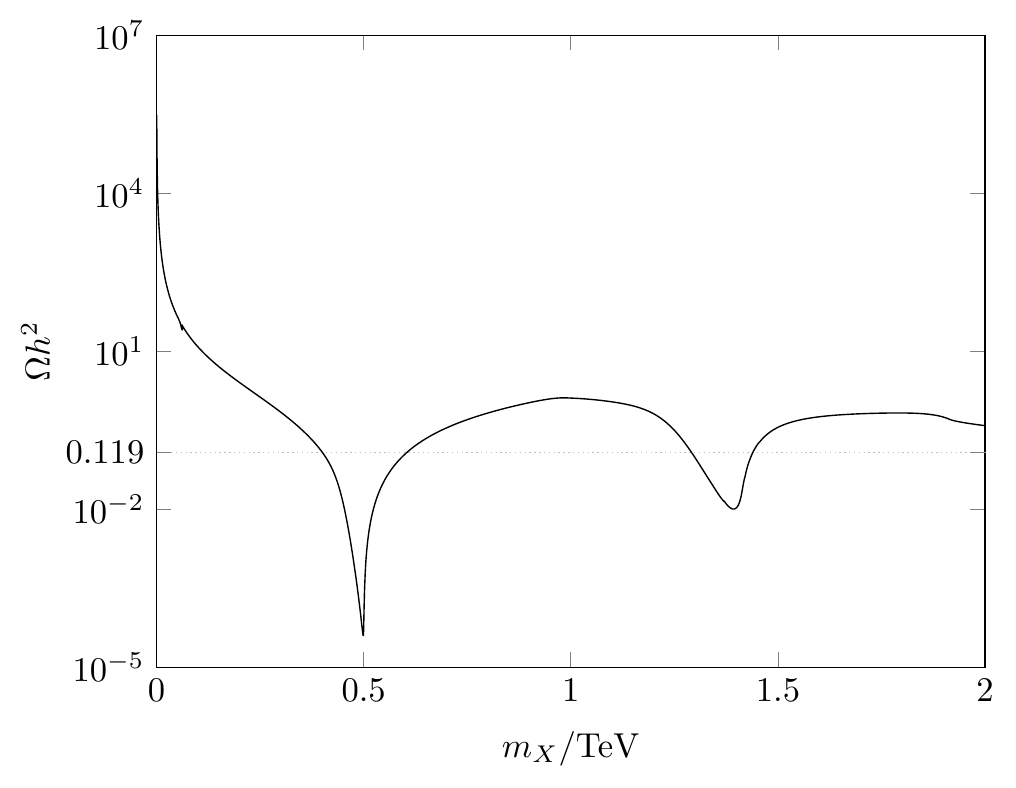}
\end{minipage}
\begin{minipage}{0.32\textwidth}
\includegraphics[width=\textwidth]{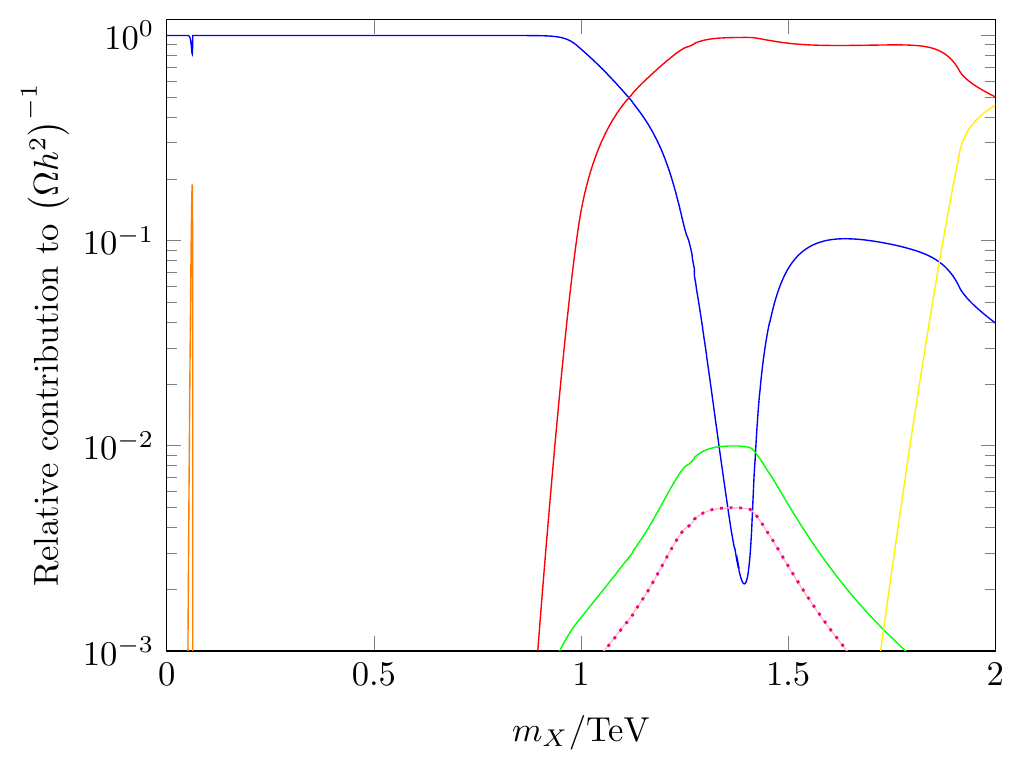}
\end{minipage}
\begin{minipage}{0.32\textwidth}
\includegraphics[width=\textwidth]{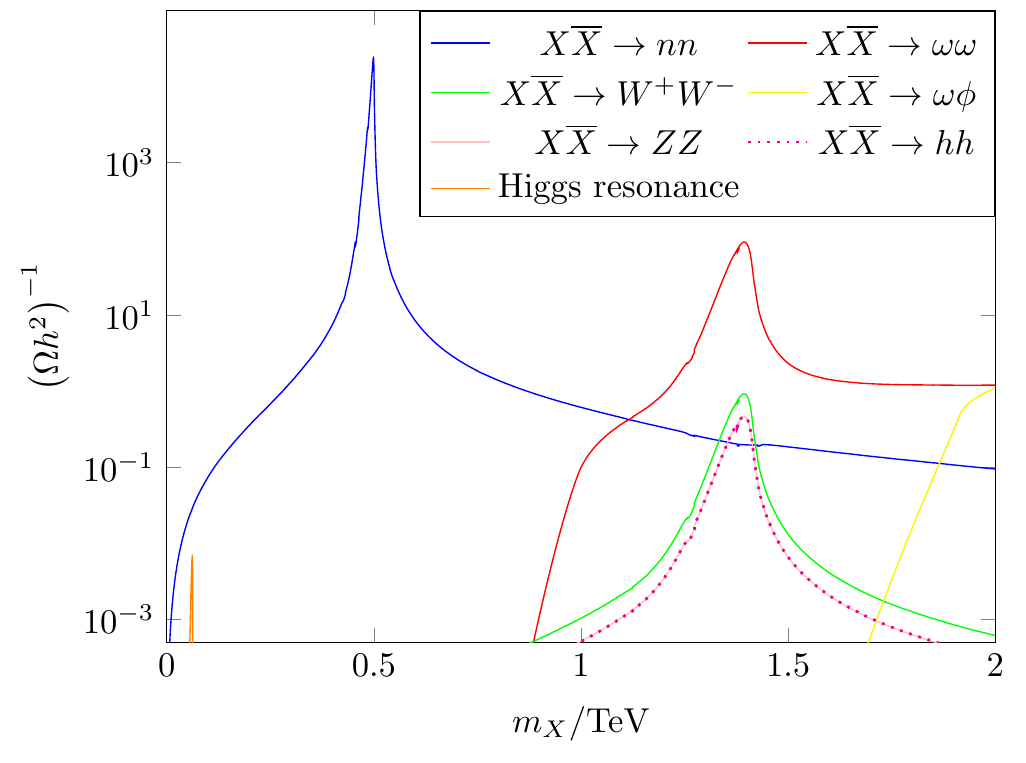}
\end{minipage}
\caption{\label{fig:abundance}{\bf Left:} DM abundance against $m_{X}$, for $v_{\phi} = \SI{2}{\tera\electronvolt}$, $m_n = \SI{100}{\kilo\electronvolt}, \lambda_{\psi} = 1, \tilde{g} = 1, \lambda_{\Phi} = 1, \lambda_{H, \Phi} = 0.1$. The horizontal dotted line indicates the observed DM abundance. The first trough corresponds to the resonant $X\bar{X}\rightarrow nn$ process with $\omega$ in the s-channel with $m_{X} \approx m_{\omega}/2$ ($m_{\omega}=\frac{\tilde{g}}{2} v_{\phi} =\SI{1}{\tera\electronvolt}$) and the second, shallower, trough corresponds to the resonant $X\bar{X}\rightarrow \omega\omega$ process with $\phi$ in the s-channel with $m_{X} \approx m_{\phi}/2$ ($m_{\phi} =  \sqrt{2\la_{\phi}} v_{\phi} \approx \SI{2.8}{\tera\electronvolt}$).
{\bf Middle (Right):} Relative (absolute) contributions of each channel to $\left(\Omega h^2\right)^{-1}$. The line labelled `Higgs Resonance' includes several channels which are only significant near the $h$ resonance at $m_X \approx m_h/2$. These are dominated by $X\overline{X} \rightarrow b\overline{b}$; the next largest contributions come from $X\overline{X} \rightarrow GG$, $X\overline{X} \rightarrow \tau^+\tau^-$ and $X\overline{X} \rightarrow c\overline{c}$. }
\end{figure}

In the left-hand panel of Figure~\ref{fig:abundance} we demonstrate how the DM abundance behaves as a function of the DM mass, $m_{X}$, with other parameters fixed at $v_{\phi} = \SI{2}{\tera\electronvolt}, m_n = \SI{100}{\kilo\electronvolt}, \lambda_{\psi} = 1, \tilde{g} = 1, \lambda_{\Phi} = 1, \lambda_{H, \Phi} = 0.1$ (unless stated otherwise these are the parameter values used in all plots in this section). The dynamics of the DM freeze-out is rather insensitive to the exact value of $m_n$ provided $m_n < m_X$ by at least a factor of 10. 

The horizontal dotted line in Figure~\ref{fig:abundance} indicates the observed DM abundance. In the middle and right-hand panel of  Figure~\ref{fig:abundance} we show the absolute and relative contributions of different DM annihilation channels to $\left(\Omega h^2\right)^{-1}$ respectively. 

For $m_X<m_{\omega}$ the dominant annihilation process is $X\bar{X}\rightarrow nn$, which proceeds via s-channel exchange of the hidden sector gauge boson, $\omega$. For $m_X>m_{\omega}$, DM annihilation into pairs of $\omega$ gauge bosons is kinematically possible and becomes the dominant channel for $m_X\sim m_{\phi}/2$ and above. 

The annihilation cross section for $X\bar{X}\rightarrow nn$ expanded in powers of relative velocity, $v$, reads
\bea\label{eqn:xsec}
\sigma(X\bar{X}\rightarrow nn)v &\approx&
\frac{\tilde{g}^4m_X^2\abs{U^{n\psi}}^4}{128\pi \left[\left(4m_X^2-m_{\omega}^2\right)^2+\Ga^2_{\omega}m^2_{\omega}\right]},
\eea 
where $\Gamma_{\omega}$  is the total decay width of $\omega$ and we have assumed\footnote{This assumption is only made in the analytic expressions for the cross section, the mass of the Portalinos is included in the numerical calculations.} $m_X\gg m_n$. For $X\bar{X}\rightarrow \omega\omega$, the cross section has the form
\bea 
\label{eq:wwfinal}
\sigma(X\bar{X}\rightarrow \omega\omega) v
 &\approx& \frac{\tilde{g}^4 \left(m_X^2-m_{\omega}^2\right)^{3/2}}{256 \pi  m_{\omega}^2 m_X \left(2 m_X^2-m_{\omega}^2\right)}+ \frac{v^2 \tilde{g}^2\;F(m_X, m_{\omega}, \tilde{g}, \la_X, \th)}{\left[\left(4m_X^2-m_{\phi}^2\right)^2+m_{\phi}^2\Ga_{\phi}^2\right]}
\eea
where $\Ga_{\phi}$ is the total decay width of  $\phi$ and $F(m_X, m_{\omega}, \tilde{g}, \la_X, \th)$ is a complicated expression detailed in Appendix \ref{app:cs}. The $\mathcal{O}(v^2)$ term in Equation~\ref{eq:wwfinal} includes the s-channel diagram with $\phi$ in the intermediate state, that, although $p$-wave, will dominate the $\sigma(X\bar{X}\rightarrow \omega\omega)$ around $m_X\sim m_{\phi}/2$. The s-wave term in Equation~\ref{eq:wwfinal} comes from a diagram with $X$ in the t-channel.

Looking more closely at the left panel in Figure~\ref{fig:abundance}, the structure of the plot is dominated by two resonances, one in each of the channels described above. The first with an on-shell $\omega$ in the s-channel appearing at $m_X\sim m_{\omega}/2$, and the second with an on-shell $\phi$ appearing at $m_X\sim m_{\phi}/2$. 

The middle and right panels of Figure~\ref{fig:abundance} demonstrate over what mass range the two processes dominantly contribute to the determination of the DM abundance. In the middle panel we plot the relative contributions of all channels with more than a $1\%$ contribution and it is clear that the hidden sector/Portalino only channels dominate. 

There are some contributions from SM model channels, all of which are enabled by the Higgs Portal via the mixing between the SM Higgs and hidden sector $\phi$. For example, contributions from the $W^+W^-$, $ZZ$, $hh$ final state channels are present due to resonant s-channel exchange of the $\phi$ field. The size of these SM channel contributions is ultimately controlled by the parameter $\la_{H,\Phi}$. Suppressing this parameter or even setting it to zero shuts off the Higgs Portal and removes the contributions from the SM channels in Figure~\ref{fig:abundance}. On one hand this may be desirable as it means the DM abundance is determined entirely by hidden sector/Portalino physics. The usually close link in freeze-out models between the annihilation process determining the abundance and the predicted signal rate in direct and indirect DM detection experiments is then decoupled. There are however still potential constraints on this model from the phenomenology of the Portalinos described in Section~\ref{section:pheno} and potential signals from indirect detection described in Section~\ref{subsection:indirect}.

Conversely, if the Higgs Portal is activated by increasing the size of $\la_{H,\Phi}$ the role of the SM states in both generating the DM abundance (mainly around the $\phi$ resonance) and in constraining the model become more important and can lead to interesting signals, for example in indirect detection signals where the DM states are annihilating to SM final states. These processes are however p-wave and therefore velocity suppressed and do not trouble current limits. 

For sufficiently large masses the process $X\bar{X}\rightarrow \phi\omega$ can play a role, with a modest dip in the abundance towards $m_X\sim \SI{2}{\tera\electronvolt}$. At smaller masses, the Higgs s-channel resonance can also contribute but only in a very narrow range, as can be seen in the middle panel of Figure~\ref{fig:abundance}. This latter channel is again only present due to the Higgs Portal and will be reduced if $\la_{H,\Phi}$ is further suppressed below the value of 0.1 used here.

\begin{figure}[t]
\centering
\begin{minipage}{0.45\textwidth}
\includegraphics[width=\textwidth]{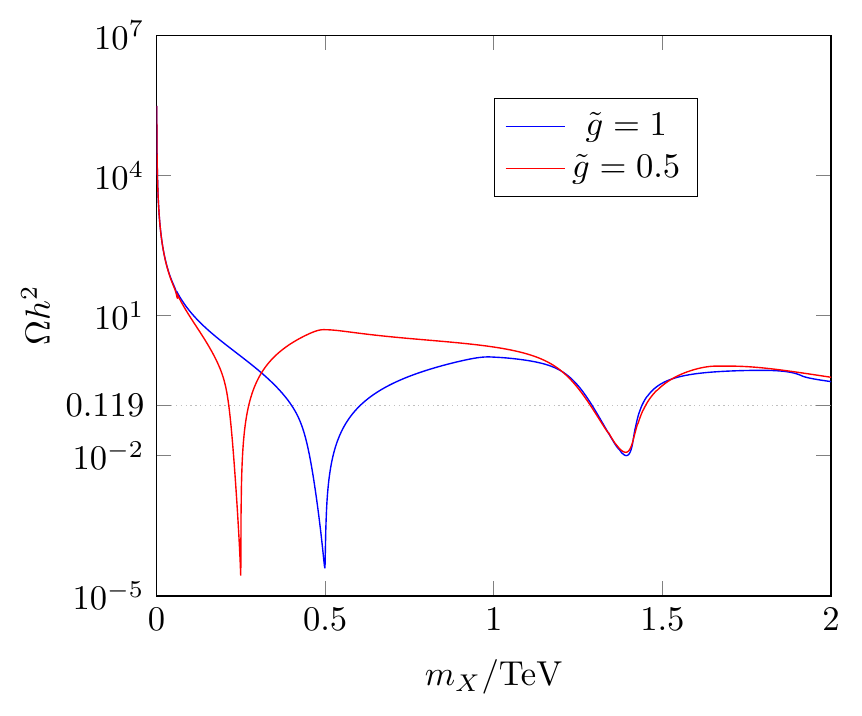}
\end{minipage}
\;\;\;\;\;\;
\begin{minipage}{0.45\textwidth}
\includegraphics[width=\textwidth]{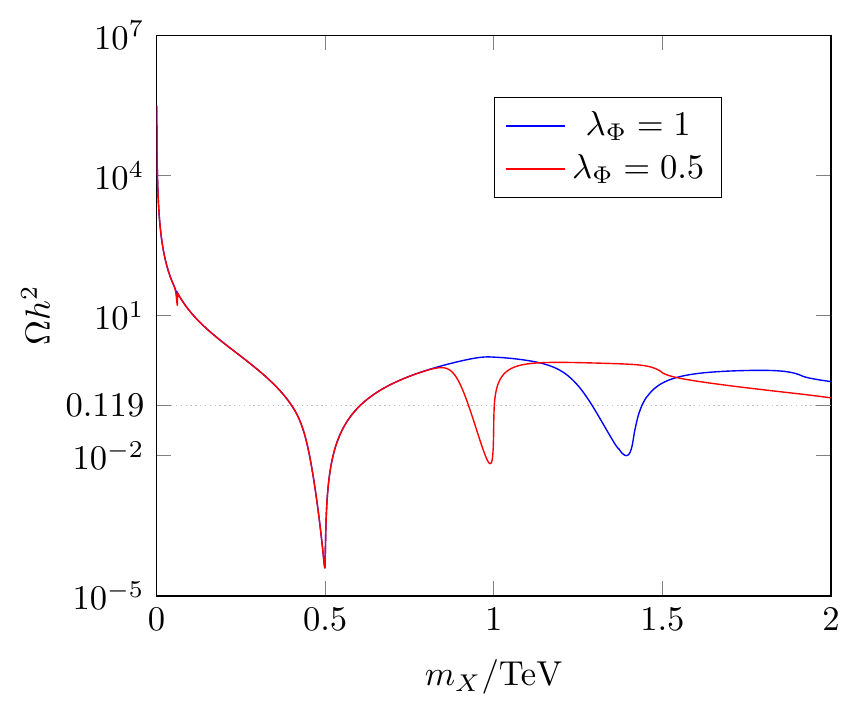}
\end{minipage}
\caption{\label{fig:reswidth} DM abundance against $m_{X}$ for different values of $\tilde{g}$ (left panel) and $\lambda_{\Phi}$ (right panel). Apart from the values of $\tilde{g}$ and $\lambda_{\Phi}$ indicated in the plots, both panels display results for parameter values as stated for the left panel of Figure~\ref{fig:abundance}. The effect of varying the values of both these parameters is seen in the position and shape of the troughs in the DM abundance (see text for details).} 
\end{figure}

In Figure~\ref{fig:reswidth}, we demonstrate the dependence of the DM abundance on $\tilde{g}$ (left panel) and $\lambda_{\Phi}$ (right panel). In particular, the way in which these parameters determine the position and shape of the troughs in the abundance. With reference to the left panel, $\tilde{g}$ modifies the abundance in three ways. For $m_X< \SI{1.25}{\tera\electronvolt}$, the process $X\bar{X}\rightarrow nn$ dominates the determination of the abundance. With the cross section for this process going as $\sim \tilde{g}^4$, reducing the value of $\tilde{g}$ increases the abundance, which can be seen in the left panel of Figure~\ref{fig:reswidth}. 

A second variation arises due to the fact that the value of $\tilde{g}$ determines the mass of the vector boson $\omega$ for fixed $v_{\phi}$ and hence shifts the position of the resonance in $m_X$ and in turn shifts where the tough appears in the abundance. Decreasing the value of $\tilde{g}$ therefore shifts the trough in the abundance to lower DM masses. 

Finally, the width of the trough/resonance depends on $\tilde{g}$ via the decay rate of $\omega$. A smaller value of $\tilde{g}$ decreases $\Gamma_{\omega}$ producing a more narrow trough/resonance. The second trough remains largely unchanged. 

In the right panel of Figure~\ref{fig:reswidth}, the dependence of the DM abundance on $\lambda_{\Phi}$ is demonstrated. The first trough is unchanged as this is dominantly determined by the $nn$ final state channel, but the decrease in $\lambda_{\Phi}$ shifts the second trough to lower $m_X$ due to the decrease in $m_{\phi}$. The width of the $\phi$ resonance/trough is narrower for smaller $\la_X$. 

We note that the modest dip in the abundance at large $m_X$ is no longer visible in the left panel of Figure~\ref{fig:reswidth} when $\tilde{g}$ is decreased. The reason is that the $X\bar{X}\rightarrow \omega\omega$ annihilation process will dominate in this mass range due to having a dominant contribution that goes like $\sim\tilde{g}^2\la_X$ compared with the leading contribution for the $\omega\phi$ channel, which goes like $\sim \tilde{g}^4$. In the left panel of Figure~\ref{fig:reswidth} however, the dip is clearly visible and appears at a lower value of $m_X$ for $\lambda_{\Phi}=0.5$ owing to the reduced value of $m_{\phi}$. 

Finally, we summarise the dependence on the remaining free parameters. For fixed $v_{\phi}$ the DM abundance doesn't depend on $\lambda_{\nu}$, $\lambda^{\psi}$ or $m_n$, as can be seen from Equations~\ref{eqn:xsec} and ~\ref{eq:wwfinal}. There is a degeneracy in these parameters whereby a change in one can be compensated by another with no effect on the DM abundance. In Section~\ref{section:pheno}, however, we show that there are constraints on the Portalino that constrain these parameters of the model. 

For fixed $\tilde{g}, \la_X, \la_{\psi}$, increasing $v_{\phi}$ increases the masses of the hidden sector states. For DM masses around the $\omega$ resonance the correct abundance can still be achieved up to $v_{\phi}\sim 100\;$TeV, whereas the correct abundance for DM masses around the $\phi$ resonance can be achieved up to $v_{\phi}\sim 7\;$TeV. However, in both these extreme cases this is only possible if we are precisely on resonance. Given the assumption that $m_n \ll v_{\phi}$ from Section~\ref{section:model} the viability of the DM model limits the maximum mass of $m_n$ to be $\sim$TeV, the precise limit depending on the mass of the DM state and the degree of tuning to the resonance one can tolerate. 

In summary, we have shown that it is relatively straightforward to reconstruct the correct DM abundance in this model with the $X$ states freezing-out dominantly via the annihilation channels: $XX\rightarrow nn$ and  $XX\rightarrow \omega\omega$. There is an important question however about the fate of the Portalino, $n$, states. It is expected that there is a significant number density of these states left after the DM states have frozen-out and all other dark sector states have decayed. The Portalino states are unstable with potentiality long lifetimes and may disrupt, for example, Big Bang Nucleosynthesis (BBN) or the Cosmic Microwave Background Radiation (CMBR) as they decay to SM particles. Constraints coming from the Portalino phenomenology are discussed in the Section~\ref{section:pheno}. They will also play an important role in indirect detection as discussed in the Section~\ref{subsection:indirect}.

\subsection{Direct Detection}
\label{subsection:direct}

Direct DM detection signals can be generated if the Higgs portal is active, that is the parameter $\la_{H,\Phi}$ is non-zero. The dominant contribution to the direct detection signal comes from Higgs exchange with scattering cross section per nucleon approximately given by \cite{Kurylov:2003ra}
\bea
\sigma\sim \frac{m_{\rm r}^2}{2\pi}\left(\frac{\la_X \sin{2\th}}{v_hm_h^2} \right)^2\;f_p^2,
\eea
where $m_{\rm r}$ is the reduced mass of the DM-proton given by $m_{\rm r}=m_Xm_p/(m_X+m_p)$ and we have assumed that the interactions with protons and neutrons are the same with
\bea
f_p = m_p\left[\sum_{u,d,s} f_{T_q}+\frac{6}{27}f_{T_G}\right] \sim 0.30\;m_p,
\eea
where, following \cite{Ellis:2018dmb}, we have used $(f_{T_u},f_{T_d},f_{T_s},f_{T_G})=(0.018, 0.027, 0.037, 0.917)$.

Assuming a small mixing angle $\th$ and applying $\la_{\Phi}v_{\phi}^2\gg \la_Hv_h^2$ to $\tan\left( 2\th\right)$ in Equation~\ref{eq:scalarmix}, we find
\bea\label{eqn:dd}
\sigma\sim 5\times 10^{-46}\;{\rm cm}^{-2}\left(\frac{\la_{H,\Phi}}{0.1}\right)^2\left(\frac{2\;{\rm TeV}}{v_{\phi}}\right)^4\left(\frac{m_X}{2\;{\rm TeV}}\right)^2\left(\frac{1}{\la_{\Phi}}\right)^2.
\eea

This value is just below the constraint from LUX-ZEPLIN (LZ), \cite{LUX-ZEPLIN:2022qhg}, at 2\;TeV. For smaller values of $m_X$, the direct detection limit decreases linearly with decreasing mass (until around 30 GeV where it flattens off) in contrast, the predicted cross section from Equation~\ref{eqn:dd} with fixed values of $\la_{H,\Phi}$, $v_{\phi}$ and $\la_{\Phi}$ decreases with $m_X^2$. As a result, masses below 2\;TeV are allowed for $\la_{H,\Phi}$, $v_{\phi}$ and $\la_{\Phi}$ fixed at the values indicated in Equation~\ref{eqn:dd}. 

To get a more general understanding of the direct detection limits Equation~\ref{eqn:dd} can be compared to a linear approximation of the LZ bound (which holds for $m_X \gtrsim \SI{40}{\giga\electronvolt})$ and reads

\bea\label{eqn:LZ}
\sigma_{\text{max}} \sim 5.5 \times 10^{-46}\;{\rm cm}^{-2}\left(\frac{m_X}{2\;{\rm TeV}}\right).
\eea

Using this, we can write:

\bea\label{eqn:dd_upper}
m_X \lesssim \SI{2.2}{\tera\electronvolt} \left(\frac{0.1}{\la_{H,\Phi}}\right)^2\left(\frac{v_{\phi}}{2\;{\rm TeV}}\right)^4 \left(\frac{\la_{\Phi}}{1}\right)^2.
\eea

Focusing now on parameter values where the observed DM abundance is correctly reproduced in the model, it is clear from Figure~\ref{fig:abundance} that we need to be near one of the troughs corresponding to the $\omega$ or $\phi$ resonances. These occur at $m_X = m_{\omega}/2$ and $m_X = m_{\phi}/2$ respectively, or equivalently at $m_X/v_{\phi} = \tilde{g}/2$ and $m_X/v_{\phi} = \sqrt{\lambda_{\phi}/2}$. Comparing these to Equation~\ref{eqn:dd_upper}, the troughs will be allowed by direct detection limits if

\bea\label{eqn:dd_omega}
\lambda_{H, \Phi} &\lesssim& 0.15 \; \lambda_{\Phi}\; \tilde{g}^{-1/2} \left(\frac{v_{\phi}}{2\;{\rm TeV}}\right)^{\frac{3}{2}}\qquad(\omega\;\; {\rm resonance}),\\
\lambda_{H, \Phi} &\lesssim& 0.12 \; \lambda_{\Phi}^{3/4}\; \left(\frac{v_{\phi}}{2\;{\rm TeV}}\right)^{\frac{3}{2}}\qquad\;\;\;\;\,(\phi\;\; {\rm resonance}). 
\eea

In summary, direct detection can play a role in limiting the allowed parameter space, but it is always possible to suppress the predicted signal by reducing the size of $\lambda_{H, \Phi}$. Reducing this parameter has no significant impact on whether the correct abundance can be achieved. This parameter, however, cannot be arbitrarily small as it provides the interaction that keeps the dark sector in thermal equilibrium.

\section{Portalino Phenomenology}
\label{section:pheno}
The Portalino mass and dark sector masses are all related to $v_{\phi}$, as a result the scale of the Portalino mass can be linked to the dark sector masses. In particular, the mass of the DM particle $X$ can be written in terms of the Portalino mass as

\begin{align}\nonumber
m_{X} &= \lambda_X v_{\phi} =  \left(\frac{\lambda^{\nu}\lambda_{X}}{\lambda^{\psi}}\right) \left(\frac{m_n}{m_{\nu}}\right)^{\frac{1}{2}} v_h \\ 
&\approx \SI{500}{\giga\electronvolt} 
 \left(\frac{\lambda_{X}}{0.25}\right)
 \left(\frac{m_n}{\si{\giga\electronvolt}}\right)^{\frac{1}{2}} 
 \left(\frac{1}{\lambda^{\psi}}\right)  
 \left(\frac{\lambda^{\nu}}{3.6 \times 10^{-5}}\right).
\end{align}

\noindent Comparing with Figure~\ref{fig:abundance}, the observed DM abundance can be produced even in scenarios with a relatively heavy Portalino. Note though that this does require a relatively small value of $\lambda_{\nu}$. This is due to the fact that increasing the Portalino mass $m_n$ in Equation~\ref{eq:portmass} while holding $v_{\phi}$ ($m_{\psi}$) constant is only possible via decreasing the Majorana mass $m_R$. In turn, a decrease in $m_R$ necessitates a smaller value of $\lambda_{\nu}$ in order to obtain the correct neutrino mass scale in Equation~\ref{eq:lightneumass}.

\begin{figure}[t]
\centering
\includegraphics[width=10cm]{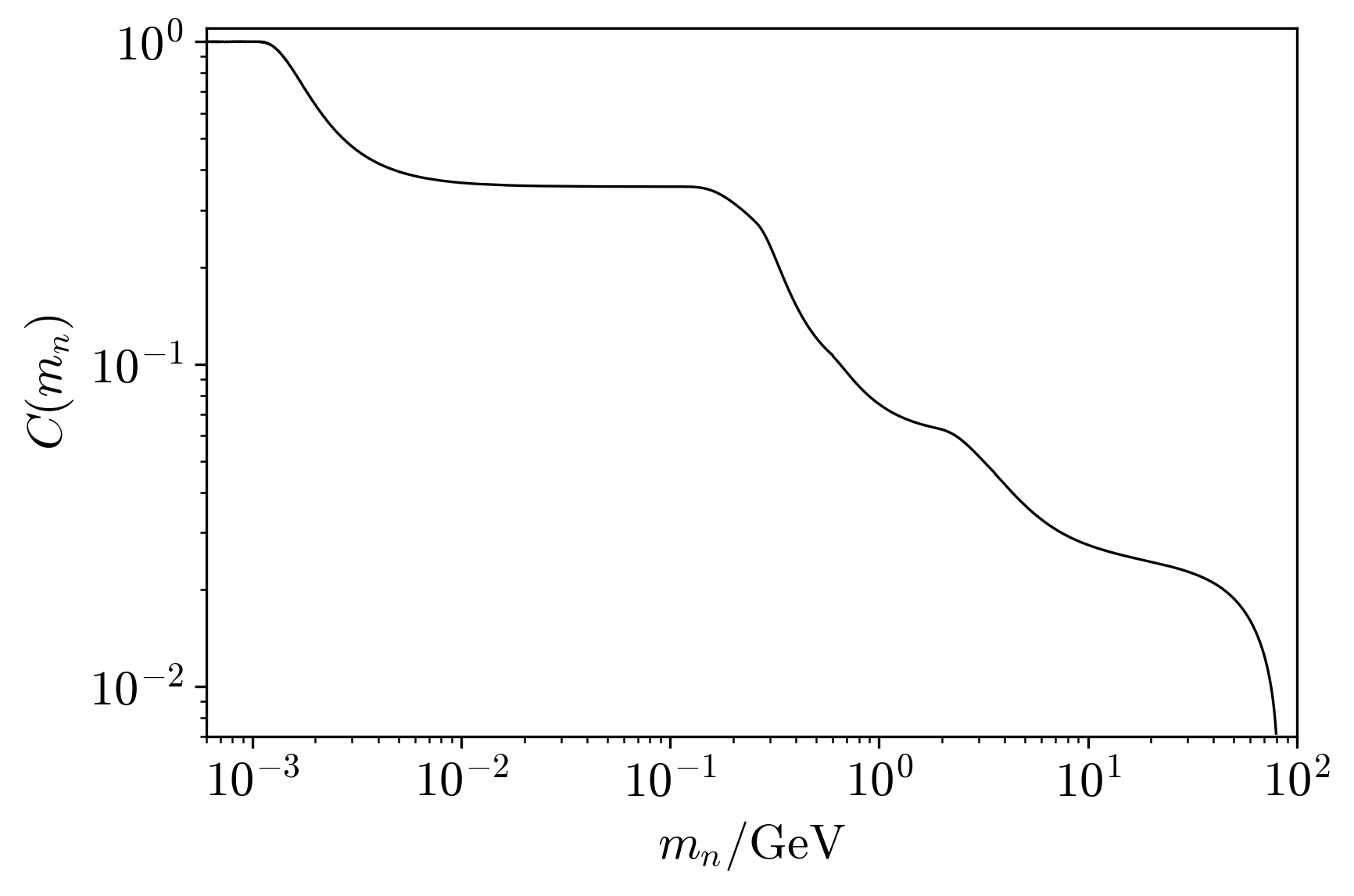}
\caption{\label{fig:cmn} The value of $C(m_n)$ from Equation~\ref{eqn:lifetime} plotted as a function of the Portalino mass $m_n$.} 
\end{figure}

Beyond the phenomenological role Portalinos play in dark matter freeze-out, indirect detection and neutrino mixing, their presence in the Early Universe may also lead to significant constraints due to their potentially very long lifetime. The decay modes for the Portalino are to SM neutrinos; a neutrino plus neutral meson; a neutrino plus charged lepton pair; or a charged lepton plus charged meson, the first two via a SM $Z$ boson and the latter two via the SM $W^{\pm}$. The mixing between the Portalino and neutrino can vary from the first order approximations described by Equation~\ref{eqn:mixings} depending on the underlying flavour parameters. To account for this in a simplified manner we include an overall scaling parameter $\eta$ in the Portalino-neutrino mixings (see Appendix~\ref{section:mixing_micro}) and neglect further details of any potential flavour structure. Within this parametrisation, the lifetime of the Portalino is given approximately by 
\begin{align}
\label{eqn:lifetime}
\tau_n \sim \SI{3}{\second} \left(\frac{1}{\eta}\right)^{2}  \left(\frac{\si{\giga\electronvolt}}{m_n}\right)^4C(m_n),
\end{align}
\noindent where $C(m_n)$ accounts for the different decay modes possible for a given mass, $m_n$. The value of $C(m_n)$ is plotted in Figure~\ref{fig:cmn} and ranges from $\sim 1$ for $m_n<2m_e$ to $\sim 0.01$ for a Portalino mass just below the $W^{\pm}$ mass. The steps down in the plot correspond to mass thresholds of Standard Models particles into which the Portalino can decay. 

The important thing to note is that these decays are cosmologically important, as any decays into the neutrino or photon sector which occur after neutrino decoupling affect the neutrino-photon temperature ratio (and hence $N_{\text{eff}}$). Furthermore there are constraints on long lived decaying particles from the primordial abundance of light elements set during BBN, \cite{Kawasaki:2017bqm}. 

This relationship between the Portalino mass and its coupling to the SM, in particular to the SM leptons, determines the Portalino decoupling temperature. Portalinos are primarily held in thermal equilibrium by processes such as $ e n \leftrightarrow e \nu$, which depends on a coupling of order $\mathcal{O} \left(m_{d}/m_{\psi}\right)$ or in terms of the physical mass states $\;\mathcal{O} \left(\sqrt{m_{\nu}/m_n}\right)$. The more massive the Portalino state is the weaker its coupling with the SM becomes and the earlier it will decouple, for example in the limit where the Portalino decouples relativistically, the rate of $ e n \leftrightarrow e \nu$ is given approximately by

\begin{equation}
\Gamma  \sim \left\{
    \begin{array}{ll}
        \eta^2 \left(\frac{m_{\nu}}{\pi^3 m_n}\right) G_F^2 T^5, & \mbox{for} \;\;T < m_Z,\\[3mm]
        \eta^2 \left(\frac{m_{\nu}}{\pi^3 m_n}\right) G_F^2 m_Z^2T^3 &  \mbox{for} \;\;T > m_Z,
    \end{array}
\right.
\end{equation}
where we have neglected the masses of the electron and neutrino and have quoted the result for decoupling temperatures above and below the mass of the $Z$ (and $W$) SM gauge boson.

By comparing these rates to the Hubble parameter we can approximately determine the decoupling temperature of the Portalinos as
\begin{align}
\label{eqn:decoupling temp}
T_{n,{\rm decouple}}   \sim \left\{
    \begin{array}{ll}
       \SI{23}{\giga\electronvolt} \left(\frac{1}{\eta}\right)^{\frac{2}{3}}\left(\frac{m_n}{\SI{1}{\giga\electronvolt}}\right)^{\frac{1}{3}}
, & \mbox{for} \;\;T_{n,{\rm decouple}} < m_Z,\\[3mm]
         \SI{165}{\giga\electronvolt} \left(\frac{0.1}{\eta}\right)^2\left(\frac{m_n}{\SI{1}{\giga\electronvolt}}\right)
 &  \mbox{for} \;\;T_{n,{\rm decouple}} > m_Z, 
    \end{array}
\right.
\end{align}
where we have set the total number of effectively massless degrees of freedom $g_*\sim 100$ and $m_\nu=0.2\;$meV.

For these example parameter values the Portalino decouples while relativistic with a significant energy density. Any Portalino decays producing neutrinos that occur after neutrino decoupling and before/during BBN or recombination would affect $N_{\text{eff}}$ during these times. Any shift in $N_{\text{eff}}$ is tightly constrained \cite{Fields:2019pfx, Aghanim:2018eyx}. In addition, sufficiently long lived Portalinos decaying into SM particles during BBN directly impact the abundance of light elements, \cite{Kawasaki:2017bqm}, further constraining the Portalino parameter space. The constraints from BBN leave us with two options: the Portalinos must decay before BBN and neutrino decoupling, or after recombination. The latter possibility can be further split into two scenarios: one in which the Portalinos decay after recombination, and another in which the Portalinos don't decay within the lifetime of the universe. These scenarios will be discussed in the following sections.

\subsection{The Heavy Portalino Case}
\label{subsection:heavy}

We first focus on the heavy Portalino case in which Portalinos decay most rapidly. For a given mass, the value of $\eta$ will determine the decoupling temperature and decay time of the Portalino and the condition that the Portalinos must decay before BBN corresponds to a minimum value of $\eta$. The parameter $\eta$ also feeds into the mixing between the Portalino and the active neutrinos (see Appendix~\ref{section:mixing} for details). Limits on the maximum size of this mixing comes from electroweak precision tests, collider searches for the direct production of Portalinos, beam dump experiments, and measurements of meson decays such that combined with the BBN constraints limits $\eta$ to a range of allowed values. The extent of this range narrows with decreasing Portalino mass and shrinks to zero at $m_n = \SI{489}{\mega\electronvolt}$. 

\begin{figure}[t!] 
   \centering
   \includegraphics[width=13cm]{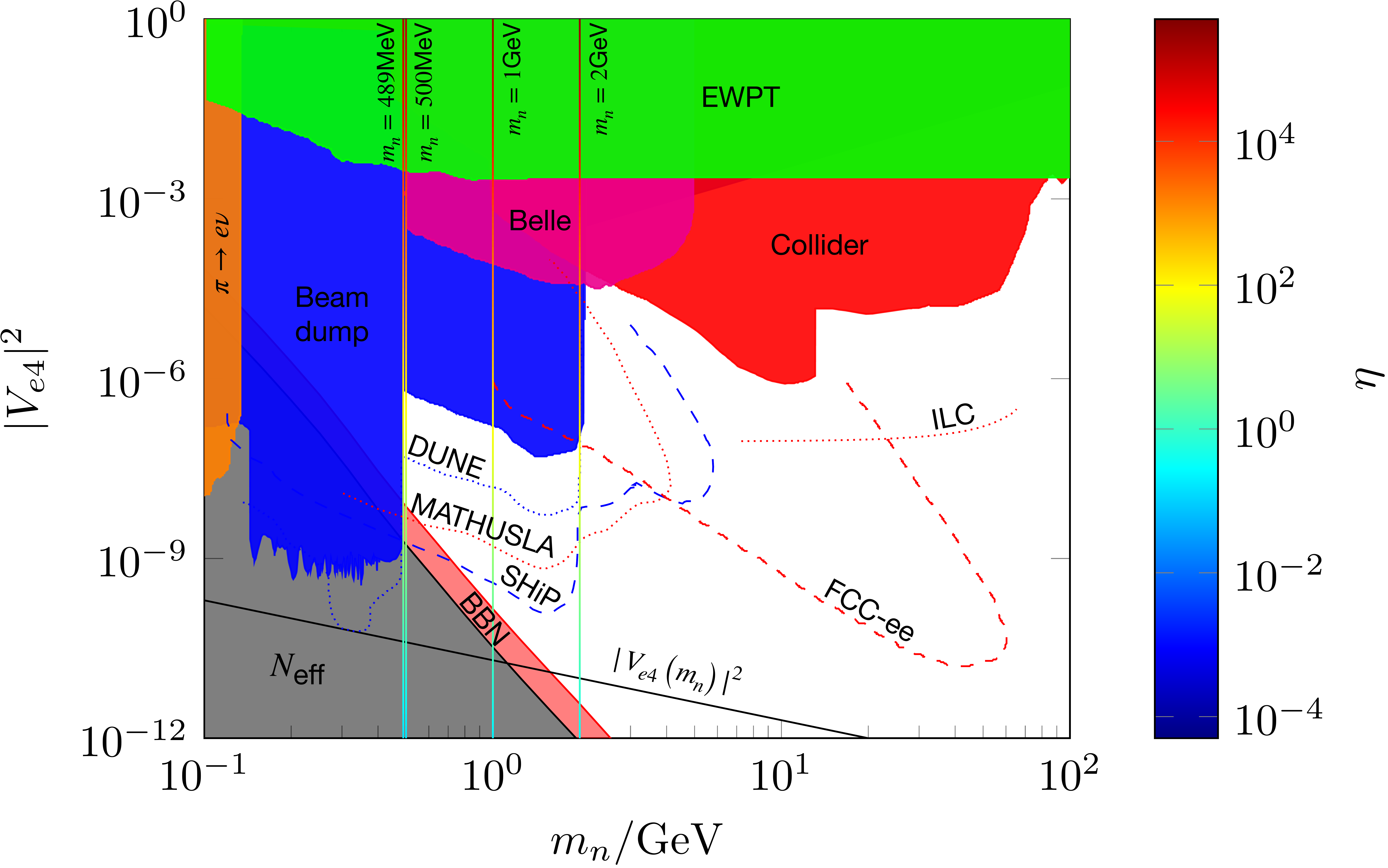}
   \caption{Current and future bounds on heavy Portalino mixing with the electron neutrino, combined with the constraint that the Portalino must decay before neutrino decoupling. See text for details of constraints such as Collider, EWPT etc.  The line labelled $| V_{e4}(m_n)|^2 $ indicates an approximate expected size of the Portalino-neutrino mixing with $\eta=1$ leaving $| V_{e4}(m_n)|^2  \sim m_{\nu}/m_n$. The four colour graded lines (for $m_n = \SI{489}{\mega\electronvolt}, \SI{500}{\mega\electronvolt}, \SI{1}{\giga\electronvolt}$ and $\SI{2}{\giga\electronvolt}$), projected into the $m_n$-$|V_{e4}|^2$ plane are lines on which the model lives as a function of $\eta$.}
   \label{fig:Exclusions}
\end{figure}

To show this, we first evaluate the combined constraints on $|V_{e 4}|^2$ for Portalino masses between $\SI{0.1}{\giga\electronvolt}$ and $\SI{100}{\giga\electronvolt}$, a summary of which is shown in Figure~\ref{fig:Exclusions}. The Portalino is not expected to mix more strongly with any particular neutrino so only constraints for $|V_{e 4}|^2$ are shown, as these are strongest. 

{\bf Electroweak precision tests:} The Portalino mixing with active neutrinos can affect several electroweak observables such as the invisible $Z$ decay width. These effects are primarily dependent on the size of the Portalino-neutrino mixing $V_{n \nu}$, but there is some mass dependence for lower values of $m_n$. Global fits of sterile neutrino mixing have been performed on electroweak precision data (taken from \cite{Deppisch:2015qwa}, which draws from \cite{delAguila:2008pw,Akhmedov:2013hec,Basso:2013jka,deBlas:2013gla,Antusch:2015mia}), and these bounds can be applied to the Portalino. This constraint is displayed in Figure~\ref{fig:Exclusions} as a green coloured region labelled `EWPT'.

{\bf Collider searches:} Portalinos can be produced directly (e.g. via $e^+ e^- \rightarrow n \nu$ or $pp \rightarrow W^* \rightarrow n l^{\pm}$) or via $Z$-boson decays. They could then decay into visible products, possibly with detectable displaced vertices. Searches for such decays have been carried out using data from LEP \cite{L3:1992xaz,DELPHI:1996qcc}, ATLAS \cite{ATLAS:2019kpx}, and CMS \cite{CMS:2018jxx, CMS:2018iaf,CMS:2022fut}. Limits have also been projected for future experiments such as MATHUSLA \cite{Curtin:2018mvb}, FCC-ee \cite{Antusch:2017pkq} and ILC \cite{Antusch:2017pkq}. These constraints are displayed in Figure~\ref{fig:Exclusions} as a red region labelled `Collider', along with projected limits for future experiments labelled `MATHUSLA', `FCC-ee' and `ILC'.

{\bf Beam dump experiments:} Portalinos with a mass of around $\SI{1}{\giga\electronvolt}$ can have significant lifetimes and so may decay at some distance from the production site. Visible decay products can be searched for in beam dump experiments with the detector positioned at a distance from the production site. Many such experiments have been carried out \cite{NOMAD:2001eyx,Ruchayskiy:2011aa,Atre:2009rg,Castro:2013jsn,Yuan:2013yba,Wang:2014lda,LHCb:2014osd}. These constraints are displayed in Figure~\ref{fig:Exclusions} as a blue region labelled `Beam dump', along with projected limits for future experiments labelled `DUNE' \cite{LBNE:2013dhi} and `SHiP' \cite{SHiP:2018xqw}. Note that CHARM and PS191 bounds have been adjusted to account for the Majorana nature of the Portalino, where the bounds are twice as strong in this case \cite{SHiP:2015gkj}. 

{\bf Meson decays:} The Portalino may take part in charged meson decays such as $X^{\pm} \rightarrow l^{\pm} n$, with a branching ratio proportional to $|V_{n \nu}|^2$. This would manifest as an additional peak in the charged meson decay spectrum. Constraints from decays such as $\pi^+ \rightarrow e^+ \nu$ are compiled in \cite{Bryman:2019ssi, Deppisch:2015qwa}, these  constraints are displayed in Figure~\ref{fig:Exclusions} labelled as `$\pi \rightarrow e \nu$'. Additionally the Belle experiment, which searched for the decay $B \rightarrow X l N$ or $B \rightarrow l N$ followed by $N \rightarrow l \pi$ (where $N$ is a sterile neutrino and X is a meson), would also place constraints on the neutrino-Portalino mixing \cite{Belle:2013ytx}. This constraint is the dark pink region labelled `Belle'.

{\bf Lepton number violation in meson decays:} The Majorana mass term violates lepton number. Hence in the Portalino model lepton number violating (LNV) processes such as $K^+ \rightarrow l^+ l^+ \pi^-$ may take place. Many searches for LNV processes have been carried out (e.g. \cite{Atre:2009rg}). However, the bounds from lepton number violation are weaker than other limits and are not shown in Figure~\ref{fig:Exclusions}.

{\bf BBN and $N_{\text{eff}}$:} Starting with the impact on $N_{\text{eff}}$ from Portalino decays into neutrinos. Neutrinos decouple from the rest of the thermal bath before electron-positron annihilation, and hence the entropy from electrons and positrons is transferred into the photons alone, raising the photon temperature relative to the neutrino temperature. In the standard case this leads to the ratio $T_{\nu} \approx \left(\frac{4}{11}\right)^{\frac{1}{3}} T_{\gamma}$. However, Portalinos can decay into neutrinos and charged leptons, so if the Portalinos decay after neutrino decoupling they will alter the neutrino-photon temperature ratio. A convenient way to parameterise this is as a constraint on the effective number of neutrino species $N_{\text{eff}}$: $\Delta N_{\text{eff}} = N_{\text{eff}} - N^{'}_{\text{eff}}$, which is constrained to be less than $0.16$ at BBN \cite{Fields:2019pfx}, and less than $0.33$ at recombination \cite{Aghanim:2018eyx} (where $N^{'}_{\text{eff}} = 3.046$ is the SM prediction \cite{deSalas:2016ztq}). 

The form of $N_{\text{eff}}$ can be defined via the relationship between the total radiation energy density  and the energy density in photons:
\begin{align}
\rho_{r} &= \rho^{'}_{\gamma} + \rho^{'}_{\nu} + \Delta_\rho \\
 &= \rho_{\gamma} \left(1 +  \frac{7}{8} N_{\text{eff}} \left(\frac{4}{11}\right)^{\frac{4}{3}}\right).
\end{align}
where $\Delta \rho$ is the energy density due to Portalino decay products, and the $'$ superscript refers to quantities ignoring any Portalino contributions. 

The size of $\Delta \rho$ depends on whether the Portalinos decouple relativistically or remain in thermal equilibrium long enough such that they freeze-out non-relativistically with a Boltzmann suppressed abundance. The latter scenario where the Portalinos freeze-out non-relativistically does not lead to a modification of $N_{\text{eff}}$ during BBN (or later) as the increased size of the coupling required to keep the Portalinos in thermal equilibrium long enough leads to a short Portalino lifetime meaning all states will have decayed well before BBN. 

For Portalinos decoupling while relativistic (and assuming they decay at a temperature $T_{n \text{,decay}} < m_n$), the energy density due to the decay products at (photon thermal bath) temperatures $T$ is given by 
\begin{align}\nonumber
\Delta \rho (T) &= m_n n_n\left(T_{n \text{,decay}}\right) \left(\frac{T}{T_{n \text{,decay}}}\right)^4 \\
&= \frac{3m_n \zeta(3)}{2 \pi^2} T_{n \text{,decouple}}^3 \left(\frac{a_{n\text{,decouple}}}{a_{n\text{,decay}}}\right)^3 \left(\frac{T}{T_{n \text{,decay}}}\right)^4,
\end{align}
where $n_n$ is the Portalino number density and the Riemann zeta function $\zeta(3)\approx 1.2$. Applying conservation of entropy we find

\begin{align}
\frac{\Delta \rho (T)}{T^4} =  \frac{3 \zeta(3)}{2 \pi^2} \frac{g_{*} (T_{n\text{,decay}})}{g_{*} (T_{n\text{,decouple}})} \frac{m_n}{T_{n\text{,decay}}}.
\end{align}

Assuming the Portalinos instantaneously decay at (thermal bath) temperature $T_{n\text{,decay}}$, they deposit energy densities $ \beta \Delta \rho$ and $(1-\beta) \Delta \rho$ into the neutrino and photon sectors respectively. The resulting change in $N_{\text{eff}}$ reads

\begin{align}
\Delta N_{\text{eff}} = \frac{15 \left( \beta \frac{8}{7} \left( \frac{11}{4} \right)^{\frac{4}{3}} - (1-\beta) N^{'}_{\text{eff}}\right)}{\pi^2 + 15(1-\beta) \frac{\Delta \rho}{(T_{n\text{,decay}})^4}}  \frac{\Delta \rho}{(T_{n\text{,decay}})^4}.
\end{align}

For simplicity in our analysis, we make the conservative choice of $\beta = 1$ when producing constraints as the Portalino abundance is so large that $\Delta N_{\text{eff}} \gg 0.16$ whenever Portalinos decay after neutrino decoupling, regardless of the value of $\beta$ (as long as $\beta \gtrsim 0.4$, below which $\Delta N_{\text{eff}}$ becomes negative). The constraint is therefore that any decay occurring after neutrino decoupling is ruled out. This constraint is the grey region labelled ``$N_{\text{eff}}$" in Figure~\ref{fig:Exclusions}.

Further to the constraints on changes to $N_{\text{eff}}$,  Portalinos decaying into SM states during BBN may directly impact the abundance of light elements. The yield of Portalinos that decouple while relativistic will be equal to the equilibrium yield, $Y_n^{\rm EQ}\sim 0.4/g_{*}(T_{\rm n, decouple})$. Even for very high decoupling temperatures, $g_{*}(T_{\rm n, decouple})$ will at most be $\sim10^2$ meaning the yield of decaying Portalinos will be large. For values of $m_n Y_n^{\rm EQ}\gtrsim10^{-8}$, the lifetime of Portalinos decaying and producing quark-antiquark pairs, for example, is restricted to less than $\sim$ 0.03 seconds, \cite{Kawasaki:2017bqm}. As with the $N_{\text{eff}}$ constraint, Portalinos decoupling when non-relativistic with a Boltzmann suppressed abundance have short lifetimes that means they will have decayed well before BBN. The constraint on relativistically decoupling Portalinos is given by the pink region labelled ``BBN'' in Figure~\ref{fig:Exclusions}, where this constraint continues ``behind'' the grey region towards the bottom left hand side of the plot. 

In Figure~\ref{fig:Exclusions} the model lives on the vertical multi coloured lines. Each line corresponds to a different Portalino mass with values 2 GeV, 1 GeV, 500 MeV and 489 MeV plotted. The colour gradient on these lines represents the changing values of $\eta$ moving from large values at the top of the figure down to small values at the bottom. As the mass of the Portalino is lowered the vertical model line moves towards the left and for $m_n=489\;$MeV the full line is completed excluded meaning that we require Portalino masses with $m_n>489\;$GeV in this scenario.

\begin{figure}[t!] 
\centering
\includegraphics[width=\textwidth]{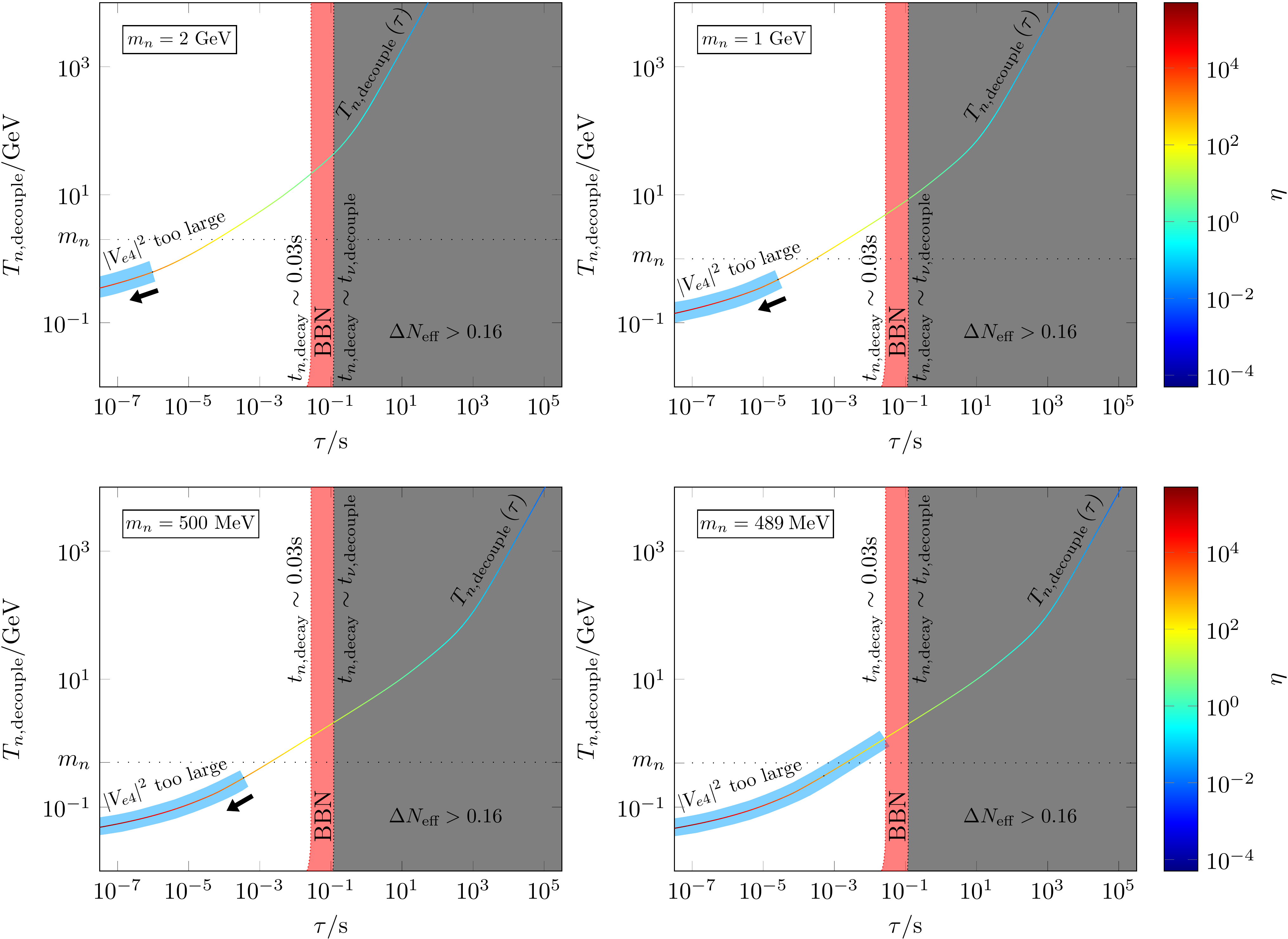}
\caption{Portalino decoupling temperature against Portalino lifetime, for a range of Portalino masses. Marked on the plot are: the decoupling temperature as a function of lifetime, $T_{n, \text{decouple}}(\tau)$, parameterised by $\eta$ (the enhancement/suppression of the Portalino-neutrino mixings); the region in pink and to the right of this is ruled out by the impact of the Portalino decays on the abundance of light elements during BBN and the grey region indicates where the Portalino decay temperature $T_{n,\text{decay}}$ is equal to or below the neutrino decoupling temperature $T_{\nu, \text{decouple}}$, within this region the Portalino decays impact on the effective relativistic degrees of freedom $N_{\text{eff}}$; and the region where the Portalino-neutrino mixing is larger than allowed by constraints is indicated by the light blue shading. The allowed region lies on the multicoloured line $T_{n, \text{decouple}}(\tau)$, between the blue shading and the pink region.}
\label{fig:Tdectau}
\end{figure}

Figure~\ref{fig:Tdectau} illustrates the same model constraints but now in terms of the Portalino decoupling temperature and lifetime. The model parameter space lies along the diagonal multicoloured line labelled $T_{n \text{,decouple}}\left(\tau\right)$. This line was calculated numerically using the exact tree-level cross section expression for the process $ e n \leftrightarrow e \nu$ using FeynCalc, \cite{Shtabovenko:2020gxv}, in order evaluate the rate $\Gamma(T, \eta) = n_n \langle \sigma v \rangle (T, \eta)$, where $n_n$ is the number density of Portlainos. 

The decoupling temperature for each value of $\eta$ was found by equating this rate to the Hubble parameter. The lifetime of the Portalino is also evaluated numerically as a function of $\eta$ to produce the multicoloured line, where the gradient of colours represents the size of $\eta$. Moving to the left on this model line the size of $\eta$ increases and as a result so does the corresponding Portalino-neutrino mixing. The blue shading around the line indicates values of $\eta$ ruled out by the constraints on $\abs{V_{e4}}^2$. 

On the right hand side of all plots in Figure~\ref{fig:Tdectau} the constraints on the Portalino lifetime stemming from $\Delta N_{\text{eff}}$ (grey region, labelled ``$\Delta N_{\rm eff}>0.16$'') and the abundance of light elements during BBN (pink region, labelled ``BBN'') are shown. The BBN constraints continue to longer lifetimes behind the grey region. 

From the fourth plot in Figure~\ref{fig:Tdectau} the limits from $\abs{V_{e4}}^2$ meet those from BBN ruling out all values of $\eta$ for Portalino masses equal to or less than 489 MeV indicating that for the heavy Portalino case we have viable parameter space for $m_n> 489$MeV for $\mathcal{O}(1)$ values of the flavour parameter $\eta$.

\subsection{Intermediate Portalino}
\label{subsection:intermediate}

Decreasing the mass of the Portalino (and/or decreasing $\eta$) allows for their decays to occur after recombination. This means that they don't affect the neutrino-photon temperature ratio at recombination, and hence they evade constraints on $N_{\text{eff}}$ at this point, potentially opening up an additional region of parameter space. However, we will show that this set-up tends to lead to an early extra period of (Portalino) matter domination, and a universe which, at the present day temperature, has an energy density and expansion rate that is too high.

Firstly, $\eta$ increases the decay rate of the Portalino. The condition that the Portalinos must decay after recombination can be recast as a condition on $\eta$:

\begin{align*}
    \textbf{Condition 1: } \tau > t_{\text{recombination}} \implies \eta \lesssim 160 \left(\frac{\SI{100}{\kilo\electronvolt}}{m_n}\right)^2.
\end{align*}

Next, we can consider the time dependence of the expansion of the universe. Similarly to the heavy case, in this scenario the Portalino tends to decouple while relativistic and with a significant number density. If $m_n \gtrsim \SI{100}{\electronvolt}$ the energy density in Portalinos comes to dominate the universe until they decay. This allows us to place a lower limit on $\eta$ given that the smaller the value of $\eta$ the longer lived the Portalinos are. An increase in the Portalino lifetime increases the length of the period of early matter domination, and leads to a lower temperature (or equivalently, a larger scale factor $a(t)$) at the point that Portalinos decay. Under the assumption that $\tau > t_{\text{recombination}}$, this doesn’t leave enough time to reach matter domination between Portalino decays and the point when the temperature of the universe reaches $T_0$ (i.e. the present day temperature). This can be seen from the following approximation for the scale factor at matter-radiation equality (where radiation includes Portalino decay products):

\begin{align}
    a_{\rm MRE} \approx 5.8 \times 10^3 \left(\frac{61.75}{g_*\left(T_{n\text{,decouple}}\right)}\right)^{\frac{2}{9}} \eta^{-\frac{4}{3}} \left(\frac{\SI{100}{\kilo\electronvolt}}{m_n}\right)^{\frac{7}{3}},
\end{align}
\noindent
where $a_{\rm MRE}>1$ would mean that the present day is radiation-dominated, with a higher expansion rate $H_0$ than observed. The condition that $a_{\rm MRE}<1$ (i.e. that matter-radiation equality is reached before the present day) can be translated into a constraint on $\eta$:

\begin{align*}
    \textbf{Condition 2: } a_{\rm MRE}<1 \implies \eta \gtrsim 670 \left(\frac{61.75}{g_*\left(T_{n\text{,decouple}}\right)}\right)^{\frac{1}{6}} \left(\frac{\SI{100}{\kilo\electronvolt}}{m_n}\right)^{\frac{7}{4}}.
\end{align*}

The final condition that must be taken into account is that $|V_{e4}|^2 < 1$ and so $\eta$ cannot be too large:

\begin{align*}
    \textbf{Condition 3: } |V_{e4}|^2 < 1 \implies \eta \lesssim 2.2 \times 10^3 \left(\frac{m_n}{\SI{100}{\kilo\electronvolt}}\right)^{\frac{1}{2}}.
\end{align*}

Conditions 1, 2 and 3 cannot be simultaneously satisfied, for any value of $m_n$. This is equivalent to the statement that the existence of a long-lived Portalino ($\tau > t_{\text{recombination}}$) which comes to dominate the universe inevitably leads to a current-day universe which is dominated by Portalino decay products (or Portalinos themselves), and is growing more quickly than we observe. Hence this scenario is ruled out.

This only leaves the possibility that the initial Portalino density is so low (via low mass and/or density) that they never come to dominate the universe, which brings us on to the next section - the light Portalino.

\subsection{Light Portalino}
\label{subsection:light}

The final possibility is a very light ($m_n \lesssim \SI{10}{\electronvolt}$) Portalino. Similarly to the above cases, DM freezes out at a temperature $T \sim \text{few hundred }\si{\giga\electronvolt}$ following this,  the Portalinos decouple (possibly long) before the QCD phase transition. Again, there is still a significant population of Portalinos after decoupling. However, unlike in either of the above cases the light Portalinos never come to dominate the energy density of the universe, and tend not to decay within the lifetime of the universe. They will however contribute $\Delta N_{\text{eff}}$ at BBN and will behave like light sterile neutrinos and will be constrained by measurements of the CMB by Planck \cite{Aghanim:2018eyx}. 

There are several other bounds for this scenario coming from the Portalino-neutrino mixing, e.g. those that arise from beta decay experiments (see for example \cite{deGouvea:2015euy}). However these bounds are far weaker than the constraints from $\Delta N_{\text{eff}}$ and Planck.

As the light Portalino will be relativistic until well after BBN, its contribution to the energy density at BBN will follow
\bea
\rho_n = \frac{7 \pi^2}{120} T_n^4 =  \frac{7 \pi^2}{120} \left(\frac{g_{*s} (T_{\nu \text{,decouple}})}{g_{*s} (T)}\right) T^4,
\eea
where $g_{*s} (T_{\nu \text{,decouple}})$ is the number of relativistic degrees of freedom at neutrino decoupling and $T$ is the temperature of the Universe. As above, $$\Delta N_{\text{eff}} = \frac{120 \rho_n}{7\pi^2T_{\nu}^4} = \left(\frac{T_n}{T_{\nu}}\right)^4 = \left(\frac{g_{*s}(T_{\nu \text{,decouple}})}{g_{*s}(T_{n \text{,decouple}})}\right)^{\frac{4}{3}}.$$

Imposing $\Delta N_{\text{eff}} < 0.16$ \cite{Fields:2019pfx} (and inserting $g_{*s}(T_{\nu \text{,decouple}}) = 43/4$), implies that $g_{*s}(T_{n \text{,decouple}}) > 42.5$, or equivalently $T_{n \text{,decouple}} \gtrsim \SI{150}{\mega\electronvolt}$ for the light Portalino. Comparing this to Equation~\ref{eqn:decoupling temp} this implies 
\bea\label{eq:lightportcon}
\eta \lesssim 0.042 \left(m_n/\SI{10}{\electronvolt}\right)^{\frac{1}{2}}.
\eea

\begin{figure}[t!] 
   \centering
   \includegraphics[width=10cm]{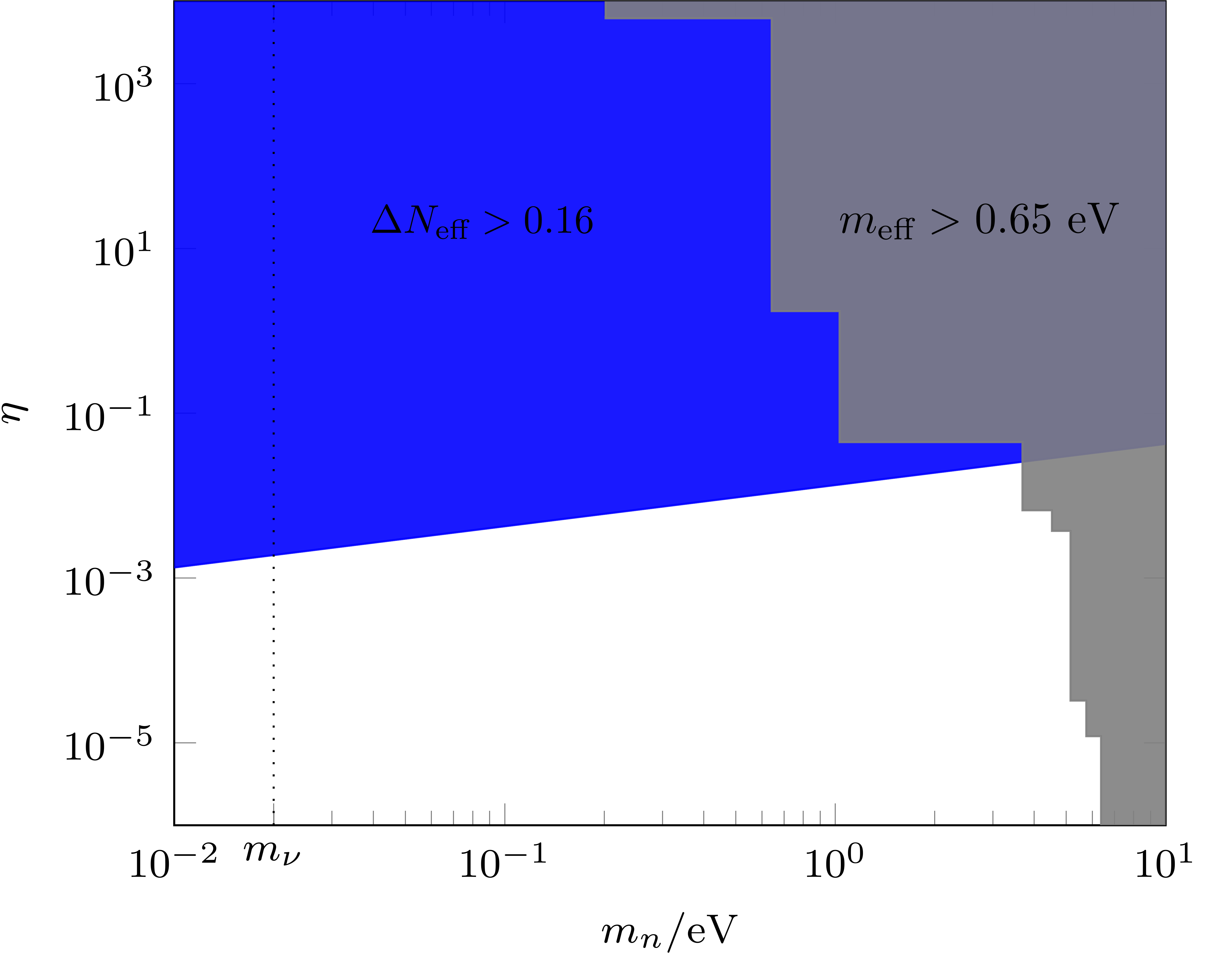}
   \caption{Constraints on the light Portalino scenario with the blue region ruled out by constraints on $\Delta N_{\rm eff}$ at BBN and the grey region ruled out by Planck's analysis of the CMB anisotropies \cite{Aghanim:2018eyx}. The vertical dashed line indicates the scale of neutrino masses and is included to highlight the limitations of the model assumption that the Portalino is more massive than the neutrinos.}
   \label{fig:light_exclusions}
\end{figure}

The second constraint on this scenario comes from Planck's determination of cosmological parameters from measurements of the CMB anisotropies, which combines data from temperature and polarisation maps with lensing and Baryon Acoustic Oscillation (BAO) measurements. In particular, the light ($m_n < \SI{10}{\electronvolt}$) Portalino is constrained by the Planck TT,TE,EE+lowE+lensing+BAO analysis limit on an effective sterile neutrino mass, $m_{\rm eff}$, where $m_{\text{eff}} = \Omega_{\text{sterile}} h^2 \left(\SI{94.1}{\electronvolt}\right)$, is constrained to be less than $\SI{0.65}{\electronvolt}$ \cite{Aghanim:2018eyx}. 

Applying this to our case the light Portalino abundance reads
\begin{align}
\Omega_n = \frac{4 \zeta (3) G T_0^3}{\pi H_0^2} \frac{g_{*S} (T_0)}{g_{*S} (T_{n \text{,decouple}})} m_n,
\end{align}
where $G$ is the gravitational constant. Hence this limit is almost entirely a constraint on $m_n$ alone, with a small adjustment depending on $T_{n \text{,decouple}}$. For example, if the Portalinos decouple extremely early, before top quarks,  $m_{\text{eff}} < \SI{0.65}{\electronvolt}$ translates to $m_n \lesssim \SI{6.4}{\electronvolt}$.

In Figure~\ref{fig:light_exclusions} the combined limits on the light Portalino scenario is mapped out as a function of $\eta$ and $m_n$ with the coloured regions ruled out. The blue region is ruled out due to the Portalino contributing too much to $\Delta N_{\rm eff}$ at BBN with the shape of the region determined by Equation~\ref{eq:lightportcon}. The grey region represents the parameter region ruled out by the Planck constraint on $m_{\rm eff}$, where the stepped shape comes from the temperature dependence of $g_{*S}$.

\subsection{Portalinos and the Indirect Detection of Dark Matter}
\label{subsection:indirect}

As outlined above there are two DM mass regions of interest corresponding to resonant annihilation processes in which the model can generate the observed DM relic abundance. For DM masses around $m_{\omega}/2$ the resonant annihilation into $nn$ pairs dominates and for masses around $m_{\phi}/2$ the resonant annihilation into $\omega\omega$ pairs dominates. The $\omega$ states decay quickly into $nn$ pairs and as a result for DM states with masses around the $\phi$ resonance, DM annihilation results in the production of four  $n$ states. 

Resonant DM annihilation into $nn$ pairs is an s-wave process and is therefore DM velocity independent. This means that the DM annihilation rate in the Galactic Centre or Dwarf Spheroidal Galaxies will be the same as that during freeze-out. In the case where the $\phi$ resonance dominates, the resonant part of the annihilation cross section to $\omega\omega$ is p-wave and with the velocity in astrophysical environments, such as the Galactic Centre, at $v\sim 10^{-3}$ (or lower in Dwarf Spheroidal Galaxies), the indirect detection signals coming from such a process will be velocity suppressed and play no role in constraining this mass region.  There is an s-wave contribution to the $\omega\omega$ channel but this is not resonant, with an annihilation rate for our parameter values well below the `thermal relic cross section' ($\sim 2\times10^{-26} {\rm cm}^3 {\rm s}^{-1}$) that is required to generate the observed relic abundance and so does not lead to constraints from indirect detection searches. 

Focusing for the rest of this Section on the s-wave DM annihilation into $nn$ pairs, the first important point is that the Portalinos produced are mono-energetic, each with an energy around the mass of the DM particle. The indirect detection phenomenology is then determined by whether or not the Portalinos decay before they reach Earth. In the case of heavy ($m_n \gtrsim \SI{489}{\mega\electronvolt}$) Portalinos produced in DM annihilations, the allowed lifetime can be up to $0.03\;$seconds (see Section~\ref{subsection:heavy}), with the Portalino travelling on average a distance, $$d\sim 10^{10}\;{\rm m} \left(\frac{m_X}{500\;{\rm GeV}}\right)\left(\frac{489\;{\rm MeV}}{m_n}\right),$$ where we have assumed the Portalinos are ultra relativistic, that is, $m_n\ll m_X$. For Portalino masses closer to the DM mass the distance travelled will be significantly shorter. Hence, for the heavy Portalino case, the Portalinos will decay before reaching Earth, even if the DM annihilation has taken place in the Sun rather than the Galactic Centre or Dwarf Spheroidal Galaxy. 

The Portalinos decay via virtual SM gauge bosons, $n \rightarrow \nu_i Z\; {\rm or}\; l^-W^+$, with the $Z$ and $W$ decaying either hadronically or leptonically leading to indirect DM signals, in particular in gamma rays. Fermi-LAT searched for gamma rays produced in DM annihilations in Milky Way Dwarf Spheroidal Galaxies, \cite{Fermi-LAT:2015att} and taking the most constrained case of DM annihilating into $\bar{b}b$ final states only they were able to rule out the thermal relic cross section for DM masses below $100\;$GeV, see e.g. Figures 1 and 2 of \cite{Fermi-LAT:2015att}. Above this mass the limit weakens and the thermal relic cross section is not constrained. Limits are also presented for DM that dominantly annihilates into $\tau^+\tau^-$ pairs, the resulting lower bound on the DM mass is only marginally lower, still of order 100 GeV. 

For $m_n>2m_b$ DM in the Portalino model does not dominantly annihilate into $\bar{b}b$ pairs, but we can use the Fermi-LAT limit to place a conservative bound on the model parameter space. For example, the lower bound on the DM mass of 100 GeV leads to $\la_X\gtrsim 0.1\; ({\rm TeV}/v_{\phi})$ and given that in order to produce the correct relic abundance we must be near the $\omega$ resonance, such that $m_X\sim m_{\omega}/2$, we have that $\tilde{g} \gtrsim 0.4\; ({\rm TeV}/v_{\phi})$. For Portalino masses between the bottom mass and the $\tau$ mass we will find similar constraints but for masses below the mass of the $\tau$, the constraints weaken considerably with the lower bound on the DM mass dropping to around 10 GeV, see for example \cite{MAGIC:2016xys} where constraints are derived for the case where the $\mu^+\mu^-$ annihilation channel dominates. Due to the Portalinos being produced on shell it is {\it their} mass that determines the spectrum of SM states produced rather than the mass of the DM, in contrast to the usual scenario of the DM annihilating directly to SM states. This offers a way to evade the indirect limits. 

The projected limits from the Cherenkov Telescope Array (CTA) will reach the thermal relic cross section (and below) for DM masses between a few 100 GeV and just over a TeV assuming DM annihilates dominantly into $W^+W^-$ or $\bar{b}b$ final states, \cite{CTA:2020hii}. These projected limits will be weakened for DM annihilating into decaying Portalinos that produce a spectrum of different SM final states or the particular the case of Portalinos with masses below the $b$-quark mass. A dedicated study of the gamma ray flux produced in this mode is required to understand both the current limits and the future constraints in detail. This includes the impact of the Portalinos travelling significant distances from the DM annihilation site before decaying into SM states. 

Beyond the signal from gamma rays, neutrinos produced in Portalino decays can also be searched for via neutrino telescopes. The DM annihilation cross section limits from IceCube \cite{IceCube:2015rnn} and ANTARES \cite{ANTARES:2022aoa} are at least an order of magnitude above the thermal relic cross section for all DM masses we consider and so do not constrain the model. For example, IceCube's most stringent limit for DM with masses between 30 GeV and 10 TeV annihilating into a pair of SM neutrinos is $\sim 4\times 10^{-24} {\rm cm}^3 {\rm s}^{-1}$, \cite{IceCube:2015rnn}.

For the lighter Portalino case ($m_n \lesssim \SI{6.4}{\electronvolt}$),  the Portalinos do not decay at all and can travel all the way to the Earth. Neutrino search experiments are potentially capable of detecting Portalinos e.g. IceCube  \cite{Abbasi:2021miz} and ANTARES \cite{ANTARES:2022aoa}, but with reduced sensitivity as the Portalinos will have a coupling to the SM suppressed by $\sqrt{m_{\nu}/m_n}$ compared with neutrino interactions and for light Portalinos of around an eV this ratio is $\sim 10^{-2}$. As with the heavy Portalino case, these Portalino states will be mono-energetic, the energy of each Portalino being equal to the DM mass, which in our model means they are produced with TeV energies. These Portalinos will produce a line in the neutrino spectrum, which provides an intriguing target for neutrino telescopes with the location of the line providing a way to measure the DM mass. As discussed above the current sensitivity of neutrino telescopes, such as IceCube, is at least an order of magnitude too weak to detect neutrinos from DM annihilations with thermal relic cross sections and with the potential suppression of Portalinos compared with SM neutrinos this signal will be challenging even with the development of the next generation of neutrino telescope. However, this does provide an exciting target to focus on given the potential for measuring both the DM mass and its annihilation cross section.

\section{Conclusions and Discussion}
\label{section:discussion}

It seems fairly natural that if a singlet right-handed neutrino does exist (as is the case in many models of neutrino mass) that it may have interactions with the hidden sector and may play a role in DM dynamics, creating a connection or portal between the SM and dark sectors. 

In this work we have expanded on the Portalino model outlined in \cite{Schmaltz:2017oov} to include neutrino masses and an expanded dark sector including a DM state. A dark $U(1)$ gauge symmetry was introduced and spontaneously broken generating the dark sector masses including for the Dirac Fermion DM state whose stability is ensured via an unbroken $Z_3$ symmetry. The observed DM abundance in this model can be generated by freeze-out via DM self-annihilations to either the Portalino states or the now massive dark sector gauge bosons. Following the freeze-out of DM, a population of Portalino states is produced. The Portalino lifetime is cosmologically relevant and as a result provides the main constraints on the properties of these states and the viability of the model. Portalinos can also potentially play an important role in the indirect detection of DM in this model. DM can annihilate into high energy mono-energetic Portalinos which in the case where they travel to the Earth can then be search for in neutrino telescopes or, if they decay, their SM decay products will lead to signals in gamma rays. Although current indirect limits do not rule these out, it is possible future experiments will née sensitive to these signals. 

We have considered three qualitatively different scenarios categorised in terms of the mass of the Portalino: an intermediate case, which is not cosmologically viable; a heavy and light case, the former with allowed parameter space for a Portalino mass $m_n \gtrsim \SI{489}{\mega\electronvolt}$ and the latter viable for $m_n \lesssim \SI{6.4}{\electronvolt}$ provided Portalinos decouple before top quarks. In the heavy Portalino case there is an upper limit of $m_n \lesssim $TeV due to the restriction that $m_n \ll v_{\phi}$, where the requirement on the successful generation of the observed DM abundance limits the maximum size of $v_{\phi}$ to the multi TeV mass scale at most. 

Throughout this work we have only considered including one Portalino (with three heavy singlet neutrinos), but we could consider models with multiple Portalinos (and/or a different number of heavy neutrinos). We can put concrete restrictions on which configurations are viable by imposing that they must give rise to at most one massless active neutrino. In the absence of specific flavour symmetries, the number of massless states is given by $n_0 = {\rm min}(0,3 - n_R + n_{p})$ where $n_R$ is the number of heavy neutrinos and $n_{p}$ is the number of Portalino states. Hence for a model with $n_p$ Portalinos the number of heavy right-handed neutrinos required is at least $2+n_p$.

The inclusion of a full neutrino flavour model was beyond the scope of this work. The details of such a flavour model will feed into the Portalino phenomenology in a more complicated way compared with the parameterisation used here in terms of $\eta$ and may lead to ways to widen the allowed parameter range found in this work. 

Another variation on what has been presented here is to remove the Majorana mass term $M_R$. In order to generate the light active neutrino masses the neutrino Yukawa couplings $\lambda_{\nu}$ would need to be small ($\sim 10^{-13}$). The structure of this model is significantly different: for example the Portalino mass isn't suppressed relative to the other dark states and the Portalino-neutrino mixing is significantly suppressed. Interestingly, if the $h-\phi$ interactions were turned off or also significantly suppressed, the Portalino-neutrino mixing could be the strongest interaction between the dark and visible sectors, and could  potentially lead to the freeze-in production \cite{McDonald:2001vt,Hall:2009bx} of the Portalino or other dark sector states.  

The Portalino can in principle provide explanations for some anomalies. Firstly, a decaying sterile state (which could be the heavy Portalino) has been proposed as a solution to short baseline anomalies \cite{Aguilar-Arevalo:2018gpe,Aguilar:2001ty, Fischer:2019fbw}. An $\si{\electronvolt}$ sterile neutrino has also been mooted as a solution to these anomalies, however the most straightforward case of an $\si{\electronvolt}$ Portalino with sufficiently strong mixing ($V_{n \nu} \sim 0.1$) with SM neutrinos would be ruled out by cosmological constraints \cite{Aguilar-Arevalo:2018gpe, Hamann:2011ge}.

\appendix

\section{Neutrino masses and mixing}
\label{section:mixing}

In this Appendix, a detailed presentation of the diagonalisation of the $(\nu_{L\alpha}, \psi, \nu_{R\alpha}$) sector is outlined. The objective is to evaluate the mass eigenstates and mixings of this seven by seven system with mass matrix given by
\[
M = 
\left(\begin{array}{cc}
	\bigzero &
	\begin{matrix}
	M_d^T\\ 
	\mathbf{M}_{\psi}^T 
	\end{matrix} \\
\begin{matrix}
M_d & \vec{M}_{\psi}
\end{matrix} & M_R
\end{array}\right), 
\]
where $M_d$ and $M_R$ are $3\times 3$ matrices and $\vec{M}_{\psi}$ is a three vector.

First it is noted that the mass matrix, $ M $, has a zero eigenvalue, $m_{\nu_1}$, with eigenvector:

\[
\vec{e}_{1} = 
N \begin{pmatrix}
- M_d^{-1} \vec{M}_{\psi} \\
1 \\
0 \\ 
0 \\ 
0 
\end{pmatrix}
\]
where $N =1/\sqrt{1 + \| M_d^{-1} \vec{M}_{\psi} \| ^2} $. We define an orthonormal basis which includes this zero eigenvector:
\[
\vec{e}_1,\;
\vec{e}_2 = \begin{pmatrix}
\vec{x}_1 \\
0 \\
0 \\ 
0 \\ 
0 
\end{pmatrix},
\vec{e}_3 = \begin{pmatrix}
\vec{x}_2 \\
0 \\
0 \\ 
0 \\ 
0 
\end{pmatrix},
\vec{e}_4 = \frac{N}{ \| M_d^{-1} \vec{M}_{\psi} \| }
\begin{pmatrix}
 M_d^{-1} \vec{M}_{\psi} \\
 \| M_d^{-1} \vec{M}_{\psi} \| ^2 \\
0 \\ 
0 \\ 
0 
\end{pmatrix},
(\vec{e}_i)_j = \delta_{ij},
\]
where $\vec{x}_{1,2} $ are chosen such that $ \vec{x}_{1, 2}^{T} M_d^{-1} \vec{M}_{\psi} = \vec{x}_1^{T} \vec{x}_2 = 0 $ and $ \| \vec{x}_{1, 2} \|^2 = 1 $. 

After rotating away the zero eigenstate, we obtain a 7$\times$7 matrix with a non-zero 6$\times$6 sub-matrix with a seesaw-type structure:
\[
P^{-1}MP = \left(\begin{array}{cc}
\bigzero_{4,4} &
\begin{matrix}
	0_{1,3} \\
	M_D
\end{matrix} \\
\begin{matrix}
	0_{3,1} & M_D^{T}
\end{matrix} & M_R
\end{array}\right),
\]
where for the sake of clarity we have indicated the dimensions of the zero matrices (e.g. $0_{n,m}$ is a $n\times m$ zero matix) and where
\[
M_D = \begin{pmatrix}
\vec{x}_1^{T} M_d^{T} \\
\vec{x}_2^{T} M_d^{T} \\
\frac{\vec{M}_{\psi}^{T}}{N \| M_d^{-1} \vec{M}_{\psi} \| }
\end{pmatrix}
\]
and
\[
P = \left(\begin{array}{cc}
\begin{matrix}
	- N  M_d^{-1} \vec{M}_{\psi}  &  \vec{x}_1 &  \vec{x}_2 & \frac{N  M_d^{-1} \vec{M}_{\psi}}{\| M_d^{-1} \vec{M}_{\psi} \|}\\[2mm]
	N & 0 & 0 & N \| M_d^{-1} \vec{M}_{\psi} \|
\end{matrix} & \bigzero_{4,3} \\[2mm]
\;\;\;\;\bigzero_{3,4} & I_3
\end{array}\right).
\].

The resulting matrix can be approximately block diagonalised, assuming the hierarchy of masses $m_d, m_{\psi} \ll m_R$:
\bea\label{eq:massmatrixblock}
Q^{-1}P^{-1}MPQ = \begin{pmatrix}
0 & 0_{1,3} & 0_{1,3} \\[1mm]
0_{3,1} &\;\;\; -M_D M_R^{-1} M_D^{T} + \mathcal{O}(M_D^3 M_R^{-2}) & 0_{3,3} \\[1mm]
0_{3,1} & 0_{3,3} &\;\; M_R + \mathcal{O}(M_D) 
\end{pmatrix}
\eea
where 
\[
Q = \begin{pmatrix}
1 & 0_{1,3}  & 0_{1,3}  \\[2mm]
0_{3,1} & \;\; I_3 - \frac{1}{2} M_D M_R^{-1} M_R^{-1} M_D^{T} + \mathcal{O}(m_D^3 M_R^{-3}) & M_D M_R^{-1} + \mathcal{O}(M_D^3 M_R^{-3}) \\[2mm]
0_{3,1} & -M_R^{-1} M_D^{T} + \mathcal{O}(M_D^3 M_R^{-3}) & \;\; I_3 - \frac{1}{2}  M_R^{-1} M_D^{T} M_D M_R^{-1} + \mathcal{O}(M_D^3 M_R^{-3})
\end{pmatrix}.
\]
We note that the eigenvalues of the $M_R$ mass matrix will correspond to the masses of the three heavy neutrino states, labelled $N_{i}$ in Section~\ref{section:model}. 

The remaining three mass eigenvalues contained within the central $3\times 3$ mass matrix block in Equation~\ref{eq:massmatrixblock} are identified as the remaining two light neutrino masses, along with the Portalino mass, $m_n$ in Section~\ref{section:model}. The explicit form of this mass matrix is given by 

\[
- M_D M_R^{-1} M_D^{T} = - \begin{pmatrix}
\vec{x}_1^{T} M_d^{T} M_R^{-1} M_d \vec{x}_1 
&\;\;\; \vec{x}_1^{T} M_d^{T} M_R^{-1} M_d \vec{x}_2 
& \;\; \frac{\vec{x}_1^{T} M_d^{T} M_R^{-1} \vec{M}_{\psi}}{N \| M_d^{-1} \vec{M}_{\psi} \|}\\[2mm]
\vec{x}_2^{T} M_d^{T} M_R^{-1} M_d \vec{x}_1 
& \;\;\; \vec{x}_2^{T} M_d^{T} M_R^{-1} M_d \vec{x}_2 
& \;\; \frac{\vec{x}_2^{T} M_d^{T} M_R^{-1} \vec{M}_{\psi}}{N \| M_d^{-1} \vec{M}_{\psi} \| } \\[2mm]
\frac{\vec{M}_{\psi}^{T} M_R^{-1} M_d \vec{x}_1}{N \| M_d^{-1} \vec{M}_{\psi} \| } 
& \frac{\vec{M}_{\psi}^{T} M_R^{-1} M_d \vec{x}_2}{N \| M_d^{-1} \vec{M}_{
\psi}\| } 
&\;\; \frac{\vec{M}_{\psi}^{T} M_R^{-1} \vec{M}_{\psi}}{N^2 \| M_d^{-1} \vec{M}_{\psi} \| ^2}
\end{pmatrix}
\].

We use the remaining freedom to choose $\vec{x}_1$ (or equivalently $\vec{x}_2$) to aid in further diagonalising. For example, choose $ \vec{x}_1 \propto \left( M_d^{T} M_R^{-1} \vec{M}_{\psi} \right) \times \left( M_d^{-1} \vec{M}_{\psi} \right)   $ (note that if this is zero then $ M_d^{T} M_R^{-1} \vec{M}_{\psi} \propto M_d^{-1} \vec{M}_{\psi} $ and hence we can choose $ \vec{x}_1 $ and $ \vec{x}_2 $ such that the 1, 3 part is already block diagonalised - so assume this isn't the case), then:

\bea\label{eq:finalmassmatrix}
- M_D M_R^{-1} M_D^{T} = - \begin{pmatrix}
c & d & 0 \\
d & e & b \\
0 & b & a
\end{pmatrix}
\eea

where
\bea\nonumber
a &=& \frac{\vec{M}_{\psi}^{T} M_R^{-1} \vec{M}_{\psi}}{N^2 \| M_d^{-1} \vec{M}_{\psi} \| ^2} = \mathcal{O} \left( \frac{m_{\psi}^2}{m_R} \right)
\\\nonumber
b &=& \frac{\vec{M}_{\psi}^{T} M_R^{-1} M_d \vec{x}_2}{N \| M_d^{-1} \vec{M}_{\psi} \| } = \mathcal{O} \left( \frac{m_d m_{\psi}}{m_R} \right)
\\\nonumber
c &=& \vec{x}_1^{T} M_d^{T} M_R^{-1} M_d \vec{x}_1 = \mathcal{O} \left( \frac{m_d^2}{m_R} \right)
\\\nonumber
d &=& \vec{x}_1^{T} M_d^{T} M_R^{-1} M_d \vec{x}_2 = \mathcal{O} \left( \frac{m_d^2}{m_R} \right)
\\\nonumber
e &=& \vec{x}_2^{T} M_d^{T} M_R^{-1} M_d \vec{x}_2 = \mathcal{O} \left( \frac{m_d^2}{m_R} \right),
\eea
where we have used the definitions in Equation~\ref{eq:flavour_defs} to write the leading order behaviour of these expressions assuming the hierarchy of masses $m_d\ll m_{\psi} \ll m_R$. Utilising this hierarchy further we can apply a rotation, $R_{24}$, to the mass matrix in Equation~\ref{eq:finalmassmatrix} such that
\[
- R_{24}^{-1}
M_D M_R^{-1} M_D^{T}
R_{24}
 = - \begin{pmatrix}
c & d & 0 \\
d & \;\; e - \frac{|b|^2}{a} & 0 \\
0 & 0 & a + \frac{|b|^2}{a}  
\end{pmatrix}
+ \mathcal{O} \left( \frac{m_d^3}{m_R m_{\psi}} \right),
\]
where 
\[
R_{24} = \begin{pmatrix}
1 & 0 & 0\\
0 & \cos \theta _{24} & \sin \theta _{24} \\
0 & \sin \theta _{24} & \cos \theta _{24} 
\end{pmatrix}, \;\;\theta_{24} = -\frac{|b|}{a} + \mathcal{O} \left( \frac{m^3_d}{m^3_{\psi}} \right).
\]
This leaves a final 2$\times$2 matrix to diagonalise. All elements are of the same order, and $ \vec{x}_2 $ is already fixed by the orthogonality constraints. A final rotation leaves the system diagonal:
\[
- R_{23}^{-1}R_{24}^{-1}
M_D M_R^{-1} M_D^{T}
R_{24}R_{23}
 = -\begin{pmatrix}
m_{\nu_2} & 0 & 0\\
0 & m_{\nu_3} & 0\\
0 & 0 & m_n
\end{pmatrix}
+ \mathcal{O} \left( \frac{m_d^3}{m_R m_{\psi}} \right)
\]
where 
\[
R_{23} = \begin{pmatrix}
\cos \theta_{23} & \sin \theta_{23} & 0\\
- \sin \theta_{23} & \cos \theta_{23} & 0\\
0 & 0 & 1
\end{pmatrix}, 
\]
\vspace{2mm}
\bea\nonumber
\cos \theta_{23} &=& \frac{\text{sign}(d)}{\sqrt{2}} \sqrt{1 - \frac{\left(c + \frac{|b|^2}{a} - e\right)}{\sqrt{\left(c + \frac{|b|^2}{a} - e \right) ^{2} + 4|d|^2}}},\\[2mm]\nonumber
\sin \theta_{23} &=&  \frac{1}{\sqrt{2}} \sqrt{1 + \frac{\left(c + \frac{|b|^2}{a} - e\right)}{\sqrt{\left( c + \frac{|b|^2}{a} - e \right) ^{2} + 4|d|^2}}}.
\eea
The three masses, $m_{\nu_{1,2}}$ and $m_n$ read
\bea \nonumber
m_{\nu_1} &=& \frac{1}{2}\left( c + e - \frac{|b|^2}{a} - \sqrt{\left( c  + \frac{|b|^2}{a}- e \right)^2 + 4|d|^2} \right) = \mathcal{O} \left(\frac{m_d^2}{m_R}\right)
\\\nonumber
m_{\nu_2} &=& \frac{1}{2}\left( c + e - \frac{|b|^2}{a} + \sqrt{\left( c + \frac{|b|^2}{a} - e \right)^2 + 4|d|^2} \right) = \mathcal{O}\left(\frac{m_d^2}{m_R}\right)
\\\nonumber
m_n &=& a + \frac{|b|^2}{a} = \mathcal{O}\left(\frac{m_{\psi}^2}{m_R}\right).
\eea

Summarising the above, the 7$\times 7$ unitary matrix that diagonalises the mass matrix, $M$, is given to leading order by

\[
V = \begin{pmatrix}
i & 0 & 0 \\
0 & iI_3 & 0 \\
0 & 0 & I_3
\end{pmatrix}PQ \begin{pmatrix}
1 & 0 & 0 \\
0 & R_{24} & 0 \\
0 & 0 & I_3
\end{pmatrix}
\begin{pmatrix}
1 & 0 & 0 \\
0 & R_{23} & 0 \\
0 & 0 & I_3 
\end{pmatrix}\begin{pmatrix}
-i & 0 & 0 \\
0 & -iI_3 & 0 \\
0 & 0 & I_3
\end{pmatrix}
\]

\[ =
\left( \begin{array}{ccc}
\begin{matrix}
	-N M_d^{-1} \vec{M}_{\psi} \\[3mm]
	N
\end{matrix} &
\begin{matrix}
 	c_{23} 	\vec{x}_1 -  s_{23} c_{24} \vec{x}_2 & s_{23} \vec{x}_1 +  c_{23} c_{24} \vec{x}_2 & \frac{c_{24} N M_d^{-1} \vec{M}_{\psi} }{\| M_d^{-1} \vec{M}_{\psi} \|}  -  s_{24} \vec{x}_2 \\[2mm]
	-  s_{23} s_{24} N \| M_d^{-1} \vec{M}_{\psi} \| & c_{23} s_{24} N \|M_d^{-1} \vec{M}_{\psi} \| & c_{24} N \| M_d^{-1} \vec{M}_{\psi} \|
\end{matrix} &
\begin{matrix}
	A \\[3mm]
	-i\vec{M}_{\psi}^{T}M_R^{-1} 
\end{matrix} \\ [6mm]
0_{3,1} & iM_R^{-1}M_D^{T}R_{24}R_{23} & \hspace{-2mm} I_3 - \frac{1}{2} M_R^{-1} M_D^{T}M_DM_R^{-1}
\end{array}\right),
\]
where
\[
A = -i\frac{\left( M_d^{-1}\vec{M}_{\psi} \right)\left( M_R^{-1}\vec{M}_{\psi} \right)^{T}}{\|M_d^{-1}\vec{M}_{\psi}\|^2} - i\left( \vec{x}_1 \right)\left( M_R^{-1}M_d\vec{x}_1 \right)^{T} - i\left( \vec{x}_2 \right)\left( M_R^{-1}M_d\vec{x}_2 \right)^{T}
\]
and $c_{23}=\cos\th_{23}$ etc. 
The order of the terms in the mixing matrix $V$ are

\[  
V\sim
\begin{pmatrix}
\mathcal{O} \left( 1 \right) & \mathcal{O} \left( 1 \right) & \mathcal{O} \left( 1 \right) & \mathcal{O} \left( \frac{m_d}{m_{\psi}} \right) & \mathcal{O} \left( \frac{m_d}{m_R} \right) & \mathcal{O} \left( \frac{m_d}{m_R} \right) & \mathcal{O} \left( \frac{m_d}{m_R} \right) \\\mathcal{O} \left( 1 \right) & \mathcal{O} \left( 1 \right) & \mathcal{O} \left( 1 \right) & \mathcal{O} \left( \frac{m_d}{m_{\psi}} \right) & \mathcal{O} \left( \frac{m_d}{m_R} \right) & \mathcal{O} \left( \frac{m_d}{m_R} \right) & \mathcal{O} \left( \frac{m_d}{m_R} \right) \\\mathcal{O} \left( 1 \right) & \mathcal{O} \left( 1 \right) & \mathcal{O} \left( 1 \right) & \mathcal{O} \left( \frac{m_d}{m_{\psi}} \right) & \mathcal{O} \left( \frac{m_d}{m_R} \right) & \mathcal{O} \left( \frac{m_d}{m_R} \right) & \mathcal{O} \left( \frac{m_d}{m_R} \right) \\
\mathcal{O} \left( \frac{m_d}{m_{\psi}} \right)  & \mathcal{O} \left( \frac{m_d}{m_{\psi}} \right) & \mathcal{O} \left( \frac{m_d}{m_{\psi}} \right) & \mathcal{O} \left( 1 \right) & \mathcal{O} \left( \frac{m_{\psi}}{m_R} \right) & \mathcal{O} \left( \frac{m_{\psi}}{m_R} \right) & \mathcal{O} \left( \frac{m_{\psi}}{m_R} \right)  \\
0 & \mathcal{O} \left( \frac{m_d}{m_R} \right) & \mathcal{O} \left( \frac{m_d}{m_R} \right) & \mathcal{O} \left( \frac{m_{\psi}}{m_R} \right) & \mathcal{O} \left( 1 \right) & \mathcal{O} \left( \frac{m_{\psi}^2}{m^2_R} \right) & \mathcal{O} \left(  \frac{m^2_{\psi}}{m^2_R}\right)  \\
0 & \mathcal{O} \left( \frac{m_d}{m_R} \right) & \mathcal{O} \left( \frac{m_d}{m_R} \right) & \mathcal{O} \left( \frac{m_{\psi}}{m_R} \right) & \mathcal{O} \left( \frac{m^2_{\psi}}{m^2_R} \right) & \mathcal{O} \left( 1 \right) & \mathcal{O} \left( \frac{m_{\psi}}{m^2_R} \right)  \\
0 & \mathcal{O} \left( \frac{m_d}{m_R} \right) & \mathcal{O} \left( \frac{m_d}{m_R} \right) & \mathcal{O} \left( \frac{m_{\psi}}{m_R} \right) & \mathcal{O} \left(  \frac{m^2_{\psi}}{m^2_R} \right) & \mathcal{O} \left(  \frac{m^2_{\psi}}{m^2_R} \right)  & \mathcal{O} \left( 1 \right)
\end{pmatrix}.
\]

\[
\arraycolsep=12pt
V\sim
\left(
\begin{array}{ccc:c:ccc}
\qquad & \qquad &\qquad &\qquad &\qquad &\qquad &\qquad\\[3pt]
\qquad & \left(U^{(\nu\nu_l)}\right)^T & \qquad & \frac{m_d}{m_{\psi}}\vec{U}^{(n\nu_l)} &  \qquad &\frac{m_d}{m_R} \left(U^{(N\nu_l)} \right)^T & \qquad \\[3pt]
\qquad  & \qquad & \qquad & \qquad & \qquad & \qquad & \qquad\\[3pt] \hdashline
 \qquad & \frac{m_d}{m_{\psi}} \left(\vec{U}^{(\nu\psi)} \right)^T &\qquad  & U^{(n\psi)} & \qquad & \frac{m_{\psi}}{m_R} \left( \vec{U}^{(N\psi)} \right)^T & \qquad  \\[3pt] \hdashline
\qquad & \qquad & \qquad & \qquad & \qquad & \qquad & \qquad\\[3pt]
 \qquad & \frac{m_d}{m_R} \left( U^{(\nu\nu_R)} \right)^T &  \qquad & \frac{m_{\psi}}{m_{R}}\vec{U}^{(n\nu_R)} & \qquad & \left( U^{(N\nu_R)} \right)^T & \qquad  \\[3pt]
\qquad & \qquad & \qquad & \qquad & \qquad & \qquad &\qquad \\[3pt]
\end{array} \right).
\]

In Equation~\ref{eqn:mixings}, the flavour mixing was parameterised by factors of $U$. These nine factors have the following forms

\bea\nonumber
U^{(\nu\nu_l)}&=&\left(\begin{array}{c} -N\left( M_d^{-1}\vec{M}_{\psi} \right)^{T}\\  c_{23}\vec{x}_1^{T}-s_{23}c_{24}\vec{x}_2^{T} \\ s_{23}\vec{x}_1^{T}+c_{23}c_{24}\vec{x}_2^{T} \end{array}\right),\;\;\;
\frac{m_d}{m_{\psi}}\vec{U}^{(\nu\psi)}= \left(\begin{array}{c}N\\-  s_{23} s_{24} N \| M_d^{-1} \vec{M}_{\psi} \| \\ c_{23} s_{24} N\| M_d^{-1} \vec{M}_{\psi} \| \end{array}\right),\\[2mm]\nonumber
\frac{m_d}{m_R} U^{(\nu\nu_R)}&=& \left(\begin{array}{c} 0_{1,3} \\ c_{23} \vec{x}_1^{T} M_d^{T} M_R^{-1} + s_{23} c_{24} \vec{x}_2^{T} M_d^{T} M_R^{-1} + s_{23} s_{24} \frac{M_R^{-1} \vec{M}_{\psi}}{N \|M_d^{-1}\vec{M}_{\psi}\|} \\ -s_{23} \vec{x}_1^{T} M_d^{T} M_R^{-1} + c_{23} c_{24} \vec{x}_2^{T} M_d^{T} M_R^{-1} + c_{23} s_{24} \frac{M_R^{-1} \vec{M}_{\psi}}{N \|M_d^{-1}\vec{M}_{\psi}\|}
\end{array}\right),\\[2mm]\nonumber
\frac{m_d}{m_{\psi}}\vec{U}^{(n\nu_l)}&=&\frac{c_{24} N M_d^{-1} \vec{M}_{\psi} }{\| M_d^{-1} \vec{M}_{\psi} \|}  -  s_{24} \vec{x}_2,\;\;\;\;\;
U^{(n\psi)}=c_{24} N \| M_d^{-1} \vec{M}_{\psi} \|, 
\\\nonumber
\frac{m_{\psi}}{m_{R}}\vec{U}^{(n\nu_R)} &=& s_{24} \frac{M_R^{-1} \vec{M}_{\psi}}{N \|M_d^{-1}\vec{M}_{\psi}\|}, \\[2mm]\nonumber
\frac{m_d}{m_R}U^{(N\nu_l)}&=& 
-i\frac{\left( M_R^{-1}\vec{M}_{\psi} \right)\left( M_d^{-1}\vec{M}_{\psi} \right)^{T}}{\|M_d^{-1}\vec{M}_{\psi}\|^2} - i\left( M_R^{-1}M_d\vec{x}_1 \right)\left( \vec{x}_1 \right)^{T} - i\left( M_R^{-1}M_d\vec{x}_2 \right)\left( \vec{x}_2 \right)^{T},
\\[2mm]\nonumber
\frac{m_{\psi}}{m_R} \vec{U}^{(N\psi)}&=&-iM_R^{-1}\vec{M}_{\psi},
\;\;\;\;U^{(N\nu_R)} = I_3 - \frac{1}{2}M_R^{-1}M_D^TM_DM_R^{-1}
\eea.

The PMNS matrix, once again assuming there is no contribution from the charged lepton sector, is then

\[
\begin{pmatrix}
\nu_e \\
\nu_{\mu} \\
\nu_{\tau} \\
\psi 
\end{pmatrix}
= V_{\rm PMNS}^{4\times 4}
\begin{pmatrix}
\nu_1 \\
\nu_2 \\
\nu_3 \\
n 
\end{pmatrix}
+ 
\begin{pmatrix}
\mathcal{O} \left( \frac{m_n}{M_R} \right) \\
\mathcal{O} \left( \frac{m_d}{M_R} \right) \\
\mathcal{O} \left( \frac{m_d}{M_R} \right) \\
\mathcal{O} \left( \frac{m_d}{M_R} \right)
\end{pmatrix},
\]
where 
\begin{multline} \nonumber
V_{\rm PMNS}^{4\times 4} = \\
\begin{pmatrix}
-N M_d^{-1} \vec{m}_n & \cos \theta_{23} \vec{x}_1 - \sin\theta_{23}\cos\theta_{24}\vec{x}_2 & \sin\theta_{23}\vec{x}_1 + \cos\theta_{23}\cos\theta_{24}\vec{x}_2 & \frac{ \cos \theta_{24} N M_d^{-1} \vec{m}_n }{\|M_d^{-1} \vec{m}_n \|} - \sin \theta_{24} \vec{x}_2 \\
N & - \sin \theta_{23} \sin \theta_{24} N \|M_d^{-1}\vec{m}_n \| & \cos\theta_{23}\sin\theta_{24}N\|M_d^{-1}\vec{m}_n\| & \cos \theta_{24} N \|M_d^{-1} \vec{m}_n \|
\end{pmatrix}
\end{multline}
\[
= \begin{pmatrix}
\mathcal{O} \left( 1 \right) & \mathcal{O} \left( 1 \right) & \mathcal{O} \left( 1 \right) & \mathcal{O} \left( \frac{m_d}{m_n} \right) \\
\mathcal{O} \left( 1 \right) & \mathcal{O} \left( 1 \right) & \mathcal{O} \left( 1 \right) & \mathcal{O} \left( \frac{m_d}{m_n} \right) \\
\mathcal{O} \left( 1 \right) & \mathcal{O} \left( 1 \right) & \mathcal{O} \left( 1 \right) & \mathcal{O} \left( \frac{m_d}{m_n} \right) \\
\mathcal{O} \left( \frac{m_d}{m_n} \right) & \mathcal{O} \left( \frac{m_d}{m_n} \right) & \mathcal{O} \left( \frac{m_d}{m_n} \right) & \mathcal{O} \left( 1 \right)
\end{pmatrix}.
\]

\section{Full Lagrangian in mass eigenbasis}
\label{section:lagrangian}

In this appendix we detail the dominant contributions to interactions in the mass eigenbasis Lagrangian. For some components more than one term is included if the dominant contribution depends on relative sizes of couplings. The full Lagrangian in the mass eignestate basis reads

\beq\nonumber
\Lagr=\Lagr_{\rm matter-scalar}+\Lagr_{\rm gauge-matter}+\Lagr_{h-\phi},
\eeq
where 
\begin{align}
\Lagr_{\rm matter-scalar} \supset  
&- \frac{m_d}{m_R} \cos \theta \left( \frac{U_{\alpha i}^{(\nu\nu_l)} \lambda^{\nu}_{\alpha \beta} U_{\beta j}^{(\nu\nu_R)} }{2} \right) \overline{\nu}_i \nu_j h   \nonumber
\\
&- \left[ \frac{m_d}{m_R} \cos \theta \left( \frac{U_{\alpha}^{(n\nu_l)} \lambda^{\nu}_{\alpha \beta} U_{\beta }^{(n\nu_R)} }{2} \right)
 + \frac{ m_{\psi} }{m_R} \sin \theta \left( \frac{ U^{(n\psi)} \lambda^{\psi}_{\alpha} U_{\al}^{(n\nu_R)} }{2} \right) \right] \overline{n} n h  \nonumber
\\
&- \bigg[ \frac{m_d}{m_R} \cos \theta \left( \frac{U_{\alpha i}^{(N\nu_l) *} \lambda^{\nu *}_{\alpha \beta} U_{\beta j}^{(N\nu_R)*} }{2} \right) \nonumber
+ \frac{ m_{\psi} }{m_R} \sin \theta \left( \frac{U_{i}^{(N\psi) *} \lambda^{\psi *}_{\alpha} U_{\alpha j}^{(N\nu_R)*} }{2} \right) \bigg] \overline{N}_i N_j h  
\\
&- \frac{m_{\psi}}{m_R} \cos \theta \left( \frac{U_{\alpha i}^{(\nu\nu_l)} \lambda^{\nu}_{\alpha \beta} U_{\beta }^{(n\nu_R)} }{2} \right) \overline{\nu_i} n h
- \cos \theta \left( \frac{U_{\alpha i}^{(\nu\nu_l)} \lambda^{\nu}_{\alpha \beta} U_{\beta j}^{(N\nu_R)} }{2} \right) \overline{\nu}_i N_j h  \nonumber
\\
&- \left[ \frac{m_d}{m_{\psi}} \cos \theta \left( \frac{U_{\alpha}^{(n\nu_l)} \lambda^{\nu}_{\alpha \beta} U_{\beta i}^{(N\nu_R)} }{2} \right)
+ \sin \theta \left( \frac{U^{(n\psi)} \lambda^{\psi}_{\alpha} U_{\alpha i}^{(N\nu_R)} }{2} \right) \right] \overline{n} N_i h  \nonumber
\\
&- \bigg[\frac{m_d^2}{m_R m_{\psi}} \cos \theta \left( \frac{U_{i}^{(\nu\psi)} \lambda^{\psi}_{\alpha} U_{\alpha j}^{(\nu\nu_R)} }{2} \right) \nonumber
- \frac{m_d}{m_R} \sin \theta \left( \frac{U_{\alpha i}^{(\nu\nu_l)} \lambda^{\nu}_{\alpha \beta} U_{\beta j}^{(\nu\nu_R)} }{2} \right) \bigg] \overline{\nu}_i \nu_j \phi 
\\
&- \frac{ m_{\psi} }{m_R} \cos \theta \left( \frac{U^{(n\psi)} \lambda^{\psi}_{\alpha} U_{\alpha }^{(n\nu_R)} }{2} \right) \overline{n} n \phi  \nonumber
- \frac{ m_{\psi} }{m_R} \cos \theta \left( \frac{U_{i}^{(N\psi) *} \lambda^{\psi *}_{\alpha} U_{\alpha j}^{(N\nu_R)*} }{2} \right) \overline{N}_i N_j \phi  
\\
&- \bigg[\frac{m_d}{m_R} \cos \theta \left( \frac{U_{i}^{(\nu\psi)} \lambda^{\psi}_{\alpha} U_{\alpha}^{(n\nu_R)} }{2} \right) 
- \frac{m_d}{m_R} \cos \theta \left( \frac{U^{(n\psi)} \lambda^{\psi}_{\alpha} U_{\alpha i}^{(\nu'nu_R)} }{2} \right)  \nonumber
\\
&\qquad \qquad \qquad - \frac{m_{\psi}}{m_R} \sin \theta \left( \frac{U_{\alpha i}^{(\nu\nu_l)} \lambda^{\nu}_{\alpha \beta} U_{\beta }^{(n\nu_R)} }{2} \right) \bigg] \overline{\nu_i} n \phi  \nonumber
\\
&- \left[ \frac{m_d}{m_{\psi}} \cos \theta \left( \frac{U_{i}^{(\nu\psi)} \lambda^{\psi}_{\alpha} U_{\alpha j}^{(N\nu_R)} }{2} \right)
- \sin \theta \left( \frac{U_{\alpha i}^{(\nu\nu_l)} \lambda^{\nu}_{\alpha \beta} U_{\beta j}^{(N\nu_R)} }{2} \right) \right] \overline{\nu}_i N_j \phi   \nonumber
\\
&- \cos \theta \left(\frac{ U^{(n\psi)} \lambda^{\psi}_{\alpha} U_{\alpha i}^{(N\nu_R)} }{2}\right) \overline{n} N_i \phi + {\rm h.c.} - \lambda_{X} \cos\theta \overline{X} X \phi + \lambda_{X} \sin\theta \overline{X} X h,
\\[30pt]
\Lagr_{\rm gauge-matter} \supset &\frac{g}{\sqrt{2}} \bigg[ \left(U_{\alpha i }^{(\nu\nu_l)} \overline{\nu}_i + \frac{m_d}{m_{\psi}} U_{\alpha}^{(n\nu_l)} \overline{n} + \frac{m_d}{m_R} U_{\alpha j}^{(N\nu_l)} \overline{N}_j \right) \gamma^{\mu} \frac{1}{2} (1 - \gamma^5) e_{\alpha} W_{\mu}^+  \nonumber
\\
&\qquad \qquad \qquad + {\rm h.c. } \bigg]   \nonumber
\\
& - \frac{\sqrt{g^2 + g'^2}}{2} \bigg[ \left( U_{\alpha i }^{(\nu\nu_l)} \overline{\nu}_i + \frac{m_d}{m_{\psi}} U_{\alpha}^{(n\nu_l)} \overline{n} + \frac{m_d}{m_R} U_{\alpha j}^{(N\nu_l)} \overline{N}_j  \right) \gamma^{\mu}  \frac{1}{2} \gamma^5 \nonumber
\\
&\qquad \qquad \qquad  \left( U_{\alpha i }^{(\nu\nu_l) *} \nu_i + \frac{m_d}{m_{\psi}} U_{\alpha}^{(n\nu_l) *} n + \frac{m_d}{m_R} U_{\alpha j}^{(N\nu_l) *} N_j  \right) Z_{\mu} \bigg] \nonumber
\\
&+ \frac{\tilde{g}}{2} \bigg[ \left( \frac{m_d}{m_{\psi}} U_{ i }^{(\nu\psi)} \overline{\nu}_i + U^{(n\psi)} \overline{n} + \frac{m_{\psi}}{m_R} U_{j}^{(N\psi)} \overline{N}_j  \right) \gamma^{\mu}   \frac{1}{2} \gamma^5 \nonumber
\\
&\qquad \qquad \qquad \left( \frac{m_d}{m_{\psi}} U_{ i }^{(\nu\psi) *} \nu_i + U^{(n\psi)*} n + \frac{m_{\psi}}{m_R} U_{j}^{(N\psi) *} N_j  \right) \omega_{\mu}  \bigg] \nonumber
\\\nonumber
&- \frac{\tilde{g}}{4} \overline{X} \gamma^{\mu} \left(1 + \gamma^5 \right) X \omega_{\mu}\\ \nonumber 
&+ \left( \frac{g^2}{2} W^+_{\mu} W^{- \mu} + \left( \frac{g^2 + g'^2}{4} \right)  Z_{\mu} Z^{\mu} \right) \nonumber
\\\nonumber
& \qquad \times \left( \cos \theta v_h  h + \frac{\cos^2 \theta }{2} h^2 - \cos\theta\sin\theta h\phi -\sin\theta v_h\phi + \frac{\sin^2\theta}{2} \phi^2 \right) 
\\[2mm]\nonumber
&+ \frac{\tilde{g}^2}{4} \omega_{\mu} \omega^{\mu} \left( \cos \theta v_{\phi}  \phi + \frac{\cos^2 \theta }{2} \phi^2 + \cos\theta\sin\theta h\phi + \sin\theta v_{\phi} h + \frac{\sin^2\theta}{2} h^2 \right),
\\[15pt]\nonumber
{\rm and}\\\nonumber
\Lagr_{h-\phi} \supset &\enspace 
\left( \lambda_H v_h \cos \theta - \frac{\lambda_{H,\Phi}v_{\phi}}{2}\sin\theta\right) \cos^2 \theta h^3
+ \frac{\lambda_H}{4} \cos^4 \theta h^4 
+ \lambda_{\Phi} v_{\phi} \cos^3 \theta \phi^3 \nonumber
\\
&+ \frac{\lambda_{\Phi}}{4} \cos^4 \theta \phi^4
+ \frac{\lambda_{H,\Phi}v_{\phi}}{2}\cos^3\theta h^2 \phi \nonumber
\\
&+\left( \frac{\lambda_{H,\Phi} v_{h}}{2}  \cos \theta - 3 \lambda_{\Phi} v_{\phi} \sin \theta + \lambda_{H,\Phi} v_{\phi} \sin \theta \right) \cos^2\theta h \phi^2  \nonumber
\\
&+ \left(\lambda_H - \frac{\lambda_{H,\Phi}}{2}\right)\cos^3\theta\sin\theta h^3 \phi
+ \frac{\lambda_{H,\Phi}}{4} \cos^4 \theta h^2 \phi^2 \nonumber
\\\nonumber
&- \left(\lambda_{\Phi} - \frac{\lambda_{H,\Phi}}{2}\right)\cos^3\theta\sin\theta h\phi^3.
\end{align}

\section{Parameterisation of the mixing matrix for numerical evaluation.}
\label{section:mixing_micro}

Without a flavour model we have no guidance for what form the components (that is, the factors of $U$ in Equation~\ref{eqn:mixings}) of the $7\times7$ unitary matrix, $V$, will take. In order to evaluate the DM phenomenology we take the following assignments

\begin{align}
U^{(\nu \nu_l)} &=  \sqrt{\left(1 - \left(\frac{m_d}{m_{\psi}}\right)^2\right) \left(1 - \left(\frac{m_{\psi}}{m_R}\right)^2\right)} N_{\rm PMNS}^{\dagger}, \;\;\;
\vec{U}^{(\nu \psi)} = \frac{1}{\sqrt{3}} \left(\begin{array}{c} 1 \\ 1 \\ 1\end{array}\right) \eta, \nonumber \\
U^{(\nu \nu_R)} &= \frac{1}{\sqrt{3}} \left(\begin{array}{ccc} 1&1&1 \\ 1&1&1 \\ 1&1&1\end{array}\right), \;\;\;
\vec{U}^{(n \nu_l)} =  \frac{1}{\sqrt{3}} \left(\begin{array}{c} 1 \\ 1 \\ 1\end{array}\right) \eta, \nonumber \\
U^{(n \psi)} &= \sqrt{\left(1 - \left(\frac{m_d}{m_{\psi}}\right)^2\right) \left(1 - \left(\frac{m_{\psi}}{m_R}\right)^2\right)}, \;\;\;
\vec{U}^{(n \nu_R)} = \frac{1}{\sqrt{3}} \left(\begin{array}{c} 1 \\ 1 \\ 1\end{array}\right), \nonumber \\
U^{(N \nu_l)} &= \frac{1}{\sqrt{3}} \left(\begin{array}{ccc} 1&1&1 \\ 1&1&1 \\ 1&1&1\end{array}\right), \;\;\;
\vec{U}^{(N \psi)} = \frac{1}{\sqrt{3}} \left(\begin{array}{c} 1 \\ 1 \\ 1\end{array}\right), \;\;\;
U^{(N \nu_R)} = \sqrt{\left(1 - \left(\frac{m_d}{m_{\psi}}\right)^2\right) \left(1 - \left(\frac{m_{\psi}}{m_R}\right)^2\right)} I_3 \nonumber,
\end{align}
where $N_{\rm PMNS}$ is the experimentally observed PMNS matrix \cite{Gonzalez-Garcia:2021dve}.

\section{Dark matter annihilation cross section to hidden vectors}
\label{app:cs}
The annihilation rate for DM states into hidden sector gauge bosons reads
\begin{align}\nonumber
&\sigma(X\bar{X}\rightarrow \omega\omega) v
 = \frac{\tilde{g}^4 \left(m_X^2-m_{\omega}^2\right)^{3/2}}{256 \pi  m_{\omega}^2 m_X \left(2 m_X^2-m_{\omega}^2\right)}+ \frac{v^2 \tilde{g}^2\;F(m_X, m_{\omega}, \tilde{g}, \la_X, \th)}{\left[\left(4m_X^2-m_{\phi}^2\right)^2+m_{\phi}^2\Ga_{\phi}^2\right]}
 \\ \nonumber
 &\equiv \frac{\tilde{g}^4 \left(m_X^2-m_{\omega}^2\right)^{3/2}}{256 \pi  m_{\omega}^2 m_X \left(2 m_X^2-m_{\omega}^2\right)}+
 \frac{v^2 \tilde{g}^2\sqrt{m_X^2-m_{\omega}^2}}{6144 \pi  m_X\left[\left(4 m_X^2-m_{\phi}^2\right)^2+m_{\phi}^2 \Gamma_{\phi}^2\right] \left(m_{\omega}^3-2 m_{\omega} m_X^2\right)^4 }\\\nonumber
&\times
\left[2 K_1 \lambda_X m_{\omega} \left(m_{\omega}^2-2 m_X^2\right)^2 +18 K_2 \lambda_X^2 m_{\omega}^2 \left(m_{\omega}^2-2 m_X^2\right)^4 +\tilde{g}^2 K_4 \left(\left(m_{\phi}^2-4 m_X^2\right)^2+m_{\phi}^2 \Gamma_{\phi}^2\right) \right.  \\\nonumber &\hspace{10mm}\left.  +8 K_3\; \tilde{g} \lambda_X m_{\omega} m_X \left(m_{\phi}^2-4 m_X^2\right) \left(m_{\omega}^2-2 m_X^2\right)^2\right],
\end{align}
 where
\begin{align}\nonumber
K_1& = 3 K_2 \lambda_X (4\cos{2 \th}+\cos 4\th)\, m_{\omega} \left(m_{\omega}^2-2 m_X^2\right)^2 +4K_3\cos{2 \th}\,\tilde{g}\, m_X \left(m_{\phi}^2-4 m_X^2\right)\\\nonumber
K_2 & =4 m_X^4 -4 m_{\omega}^2 m_X^2+3 m_{\omega}^4, \\\nonumber
K_3 &= 10 m_{\omega}^4 m_X^2-24 m_{\omega}^2 m_X^4+m_{\omega}^6+16m_X^6,\\\nonumber
K_4 &= 3 m_{\omega}^8 m_X^2-84 m_{\omega}^6 m_X^4+152 m_{\omega}^4 m_X^6-80 m_{\omega}^2 m_X^8+7 m_{\omega}^{10}+32 m_X^{10}.
 \end{align}

\bibliography{port} {}
\bibliographystyle{JHEP}
\end{document}